\documentclass[twocolumn]{aastex631}
\usepackage{amsmath,esint}

\shortauthors{Farnocchia et al.}

\graphicspath{{./}{figures/}}

\begin{document}

\title{(523599) 2003~RM: The asteroid that wanted to be a comet}

\correspondingauthor{Davide Farnocchia}
\email{Davide.Farnocchia@jpl.nasa.gov}

\author[0000-0003-0774-884X]{Davide Farnocchia}
\affiliation{Jet Propulsion Laboratory, California Institute of Technology,
4800 Oak Grove Dr.,
Pasadena, CA 91109, USA}

\author[0000-0002-0726-6480]{Darryl Z. Seligman}
\affiliation{Department of Astronomy and Carl Sagan Institute, Cornell University, 122 Sciences Drive, Ithaca, NY, 14853, USA}

\author[0000-0002-5624-1888]{Mikael Granvik}
\affiliation{Asteroid Engineering Laboratory, Lule\r{a} University of Technology,
Box 848,
SE-98128 Kiruna, Sweden
}
\affiliation{Department of Physics, University of Helsinki,
PO Box 64,
FI-00014 Helsinki, Finland
}

\author[0000-0001-6952-9349]{Olivier Hainaut}
\affiliation{ESO,
Karl-Schwarzschild-Stra{\ss}e 2,
85748 Garching-bei-M\"{u}nchen, Germany
}

\author[0000-0002-2058-5670]{Karen J. Meech}
\affiliation{Institute for Astronomy, University of Hawaii,
2680 Woodlawn Dr.,
Honolulu, HI 96822, USA}

\author[0000-0001-7895-8209]{Marco Micheli}
\affiliation{ESA NEO Coordination Centre,
Largo Galileo Galilei, 1,
00044 Frascati (RM), Italy}

\author[0000-0002-0439-9341]{Robert Weryk}
\affiliation{Physics and Astronomy, The University of Western Ontario,
1151 Richmond St.,
London, ON N6A 3K7, Canada}

\author[0000-0003-3240-6497]{Steven R. Chesley}
\affiliation{Jet Propulsion Laboratory, California Institute of Technology,
4800 Oak Grove Dr.,
Pasadena, CA 91109, USA}

\author{Eric J. Christensen}
\affiliation{Lunar and Planetary Laboratory, University of Arizona,
1629 E University Blvd.,
Tucson, AZ 85721, USA
}

\author[0000-0001-8690-3507]{Detlef Koschny}
\affiliation{ESA NEO Coordination Centre,
Via Galileo Galilei, 1,
00044 Frascati (RM), Italy}
\affiliation{ESA ESTEC,
Keplerlaan 1,
2201 AZ Noordwijk, The Netherlands}
\affiliation{LRT, TU Munich,
Boltzmannstra{\ss}e 15,
85748 Garching bei M\"{u}nchen, Germany}

\author[0000-0002-4734-8878]{Jan T. Kleyna}
\affiliation{Institute for Astronomy, University of Hawaii,
2680 Woodlawn Dr.,
Honolulu, HI 96822, USA}

\author[0000-0002-4470-6043]{Daniela Lazzaro}
\affiliation{Observat\'{o}rio Nacional,
R. Gen. Jos\'{e} Cristino, 77,
20921-400 Rio de Janeiro, RJ, Brazil
}

\author[0000-0002-8132-778X]{Michael Mommert}
\affiliation{School of Computer Science, University of St. Gallen,
Rosenbergstrasse 30,
CH-9000 St. Gallen, Switzerland
}


\author[0000-0002-1341-0952]{Richard Wainscoat}
\affiliation{Institute for Astronomy, University of Hawaii,
2680 Woodlawn Dr.,
Honolulu, HI 96822, USA}



\begin{abstract}
We report a statistically significant detection of nongravitational acceleration on the sub-kilometer near-Earth asteroid (523599) 2003 RM. Due to its orbit, 2003 RM experiences favorable observing apparitions every 5 years. Thus, since its discovery, 2003 RM has been extensively tracked with ground-based optical facilities in 2003, 2008, 2013, and 2018. We find that the observed plane-of-sky  positions cannot be explained with a purely gravity-driven trajectory. Including a transverse nongravitational acceleration allows us to match all observational data, but its magnitude is inconsistent with perturbations typical of asteroids such as the Yarkovsky effect or solar radiation pressure. After ruling out that the orbital deviations are due to a close approach or collision with another asteroid,
we hypothesize that this anomalous acceleration is caused by unseen cometary outgassing. A detailed search for evidence of cometary activity with archival and deep observations from Pan-STARRS and the VLT does not reveal any detectable dust production. However, the best-fitting H$_2$O sublimation model allows for brightening due to activity consistent with the scatter of the data. We estimate the production rate required for H$_2$O outgassing to power the acceleration, and find that, assuming a diameter of 300 m, 2003 RM would require Q(H$_2$O)$\sim10^{23}$ molec/s at perihelion. We investigate the recent dynamical history of 2003 RM and find that the object most likely originated in the mid-to-outer main belt ($\sim86\%$) as opposed to from the Jupiter-family comet region ($\sim11\%$). Further observations, especially in the infrared, could shed light on the nature of this anomalous acceleration. 
\end{abstract}


\section{Introduction} \label{sec:intro}  
Modeling the trajectory of small bodies is a non-trivial problem. 
As the data quality improves and observational arcs get extended, nongravitational perturbations can become a significant consideration.
Comets are especially affected by nongravitational forces because of  sublimation and outgassing of volatiles \citep{Meech2004}. Therefore, orbit determination centers  that estimate cometary orbits such as the Jet Propulsion Laboratory\footnote{\url{https://ssd.jpl.nasa.gov}} and the Minor Planet Center\footnote{\url{https://minorplanetcenter.net}} account for nongravitational perturbations in the orbit model. Typically, these perturbations are incorporated based on the standard model of  \citet{Marsden1973}, although more sophisticated models \citep[e.g.,][]{Yeomans89,Krolikowska2004,Yeomans2004,Chesley2005} are sometimes employed.

As opposed to cometary orbits, the trajectories of 
asteroids are generally better approximated with a motion purely driven by gravitational forces. In fact, while asteroids are also subject to nongravitational perturbations such as solar radiation pressure \citep{Vokrouhlicky2000} and the Yarkovsky effect \citep{Vokrouhlicky2015_ast4}, these are orders of magnitude weaker than forces induced via outgassing. These weaker forces can still be detectable with astrometric data, especially with radar measurements \citep{Ostro2002}  and/or long observational arcs. Solar radiation pressure has been measured for a handful of small near-Earth asteroids \citep[MPEC 2008-D12,\footnote{\url{https://www.minorplanetcenter.org/mpec/K08/K08D12.html}}][]{Micheli2012,Micheli2013,Micheli2014,Mommert2014bd,Mommert2014md,Farnocchia2017TC25,Fedorets2020}, while searches for detections of the Yarkovsky effect are performed regularly \citep[e.g.,][]{Farnocchia2013_yarko,DelVigna2018,Greenberg2020} and have led to hundreds of detections.

Asteroid (523599) 2003~RM was discovered by the Near Earth Asteroid Tracking Program \citep{Pravdo1999} on 2003 September 2
(MPEC 2003-R17).\footnote{\url{https://www.minorplanetcenter.net/mpec/K03/K03R17.html}}
2003~RM has a semimajor axis $a = 2.92$ au and eccentricity $e=0.60$, which yields apsidal distances of $q \times Q = 1.17$ au $\times$ 4.68 au. The inclination is 10.9$^\circ$ relative to the ecliptic plane.
Since the initial discovery, there have been over 300 optical observations of 2003~RM  over the course of four apparitions in 2003, 2008, 2013, and 2018.
While searching for the Yarkovsky effect  among near-Earth asteroids, \citet{Chesley2016} detected a clear signal of a transverse acceleration in the motion of 2003~RM. They noted that the observed acceleration was far too large to be caused by the Yarkovsky effect, and argued that cometary activity was the most likely explanation. However, \citet{Chesley2016} also reported that closer inspection of observational images of  2003~RM by a number of near-Earth object search programs \citep{Christensen2016,McMillan2016,Wainscoat2016} did not reveal any clear evidence of  cometary activity. To date, no detection of cometary activity has been reported to the Minor Planet Center and therefore 2003~RM currently remains classified as an asteroid.

A similar puzzle affects the understanding of the cometary nature of 1I/`Oumuamua, the first interstellar object to be discovered passing through the inner Solar System (MPECs 2017-U181 and 2017-V17).\footnote{\url{https://minorplanetcenter.net//mpec/K17/K17UI1.html}}\textsuperscript{,}\footnote{\url{https://minorplanetcenter.net/mpec/K17/K17V17.html}}
While no cometary activity was detected around `Oumuamua \citep{Meech2017,Trilling2018}, its astrometric positions could only be explained with the addition of nongravitational perturbations, most plausibly due to outgassing activity \citep{Micheli2018}. Additionally, `Oumuamua had an extreme shape, a reddened reflection spectrum \citep{Meech2017} and a low incoming velocity indicating a young $<40$ Myr age \citep{Feng2018}.
Reconciling the lack of activity and the observed nongravitational perturbations remains a challenge \citep{Jewitt2022}, even though models have been proposed that could explain all the observed properties of `Oumuamua, e.g., by assuming a significant presence of molecular hydrogen ice \citep{Seligman2020}. Alternative explanations have invoked the presence of N$_2$ and CO driven activity \citep{Desch20211i,Jackson20211i,Seligman2021}  or radiation pressure \citep{Micheli2018}. 
Spin-up was detected in the object \citep{Taylor2022} consistent with outgassing torques \citep{Rafikov2018}. There is precedent for outgassing without dust activity, such as  the CO enriched active Centaur 29P/Schwassmann-Wachmann 1 \citep{Senay1994,Crovisier1995,Gunnarsson2008,Paganini2013}, which exhibited CO and dust outbursts that were not well correlated in time \citep{Wierzchos2020}.

\section{Astrometry} \label{sec:astrometry}

2003~RM is currently in a 5:1 mean motion resonance with the Earth. Subsequently, the object has apparitions visible from the Earth every five years. Since its discovery in 2003, the object has been extensively tracked with ground-based optical facilities over the course of the following four apparitions:
\begin{itemize}
    \item From 2003 September 2 to 2003 October 19: 85 observations;
    \item From 2008 June 24 to 2008 October 29: 73 observations;
    \item From 2013 May 16 to 2013 October 24: 45 observations;
    \item From 2018 March 18 to 2018 November 14: 98 observations.
\end{itemize}

The full observational dataset is available from the Minor Planet Center.\footnote{\url{https://minorplanetcenter.net/db_search/show_object?object_id=523599}}
For the majority of the existing astrometric data,  observations were weighted using the scheme presented by \citet{Veres2017}. Positional uncertainties were estimated for our own measurements:
\begin{itemize}
    \item Siding Spring Survey (station code E12)\footnote{\url{https://minorplanetcenter.net/iau/lists/ObsCodesF.html}} observations on 2008 June 24 and 25, which we remeasured and weighted at 0.5$''$;
    \item Catalina Sky Survey (station code 703) observations on 2008 September 7 (weights at 0.2$''$) and 21 (0.6$''$), 2013 August 28 (0.8$''$), 2013 September 15 (0.5$''$) and 23 (0.8$''$), which we remeasured;
    \item Pan-STARRS 1 (station code F51) observations on 2013 July 7 (0.2$''$), 2013 August 17 (0.1$''$) and 28 (0.2$''$), 2013 September 14 (0.1$''$), and 2013 October 24 (0.1$''$);
    \item OASI, Nova Itacuruba (station code Y28) observations on 2018 March 18 (0.7$''$), 2018 April 19 (0.1$''$) and 20 (0.2$''$), 2018 May 17 (0.2$''$), 2018 July 09 (0.2$''$), 15 (0.1$''$), 17 (0.1$''$), and 18 (0.1$''$), 2018 August 7 (0.1$''$), 2018 September 6 (0.1$''$) and 11 (0.2$''$);
    \item Las Campanas Observatory (station code 304), with the Magellan Baade telescope, on 2018 June 22 (0.1$''$);
    \item Lowell Discovery Telescope (station code G37) on 2018 September 4 (0.15$''$);
    \item Very Large Telescope (station code 309) with the Unit Telescope 1, observations on 2018 September 19 (0.1$''$).
\end{itemize}
We assumed a 1 s uncertainty in the reported time of observation and corrected for star catalog systematic errors using the \citet{Eggl2020} debiasing scheme.
To reject outliers, we employed the \citet{Carpino2003} algorithm.

\section{Detection of nongravitational perturbations} \label{sec:nongravs}
Our default gravitational force model configuration \citep[e.g.,][]{Farnocchia2015ast4} is based on JPL planetary ephemeris DE441 \citep{Park2021} and the 16 most massive small-body perturbers in the main belt \citep{Farnocchia2021sb441}.\footnote{\url{ftp://ssd.jpl.nasa.gov/pub/eph/small_bodies/asteroids_de441/SB441_IOM392R-21-005_perturbers.pdf}}
As was initially pointed out by \citet{Chesley2016}, this gravity-only model configuration fails to satisfactorily match the 2003~RM astrometric data.
While any given set of two consecutive apparitions can be fit, the addition of a third apparition results in unacceptably high residuals.

\begin{figure*}[ht!]
\epsscale{0.8}
\plotone{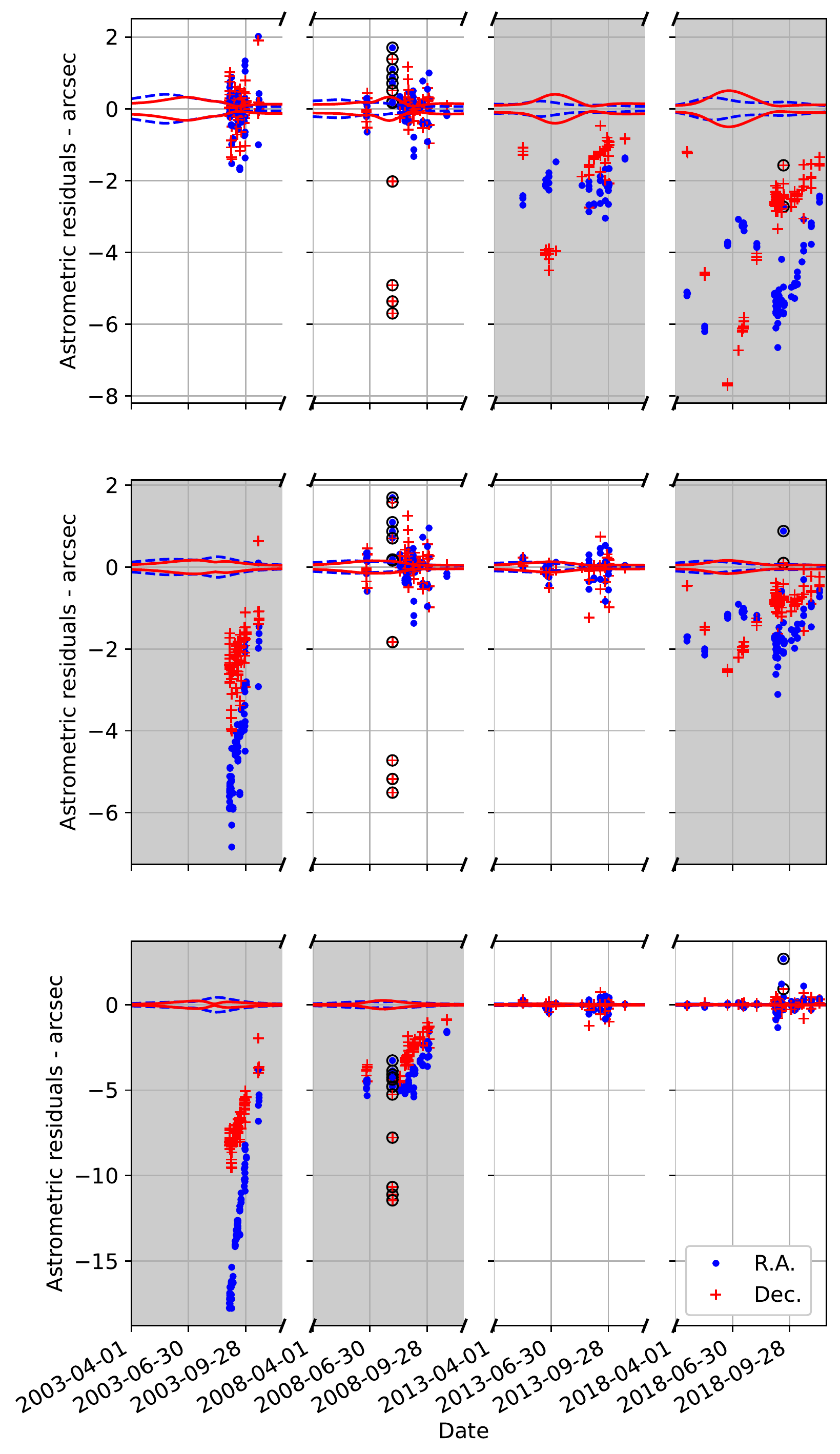}
\caption{Astrometric residuals against the best gravity-only fit to the astrometry from 2003 to 2008 (top panel), from 2008 to 2013 (middle panel), or from 2013 to 2018 (bottom panel). Dots correspond to Right Ascension (R.A.) and crosses correspond to Declination (Dec.). Outliers rejected from the fit are indicated with circles. The shaded area corresponds to observations not included in the fit. The dashed line and solid line represent the 1-$\sigma$ ephemeris uncertainties of the orbital solution in R.A. and Dec., respectively. The large residuals in the shaded areas far exceed the ephemeris prediction uncertainties and thus imply problems in the force model.}\label{fig:grav}
\end{figure*}

To illustrate this point, Figure~\ref{fig:grav} shows the astrometric residuals of the entire dataset against gravity-only solutions based on two consecutive apparitions.
Except for a handful of outliers, the residuals of the fitted astrometric observations are consistent with the assumed observational uncertainties.
However, residuals of the observations outside the two fitted apparitions are clearly too large to be explained by astrometric errors or ephemeris uncertainties.
This behavior is \textit{not} caused by only a handful of isolated observations. Instead, every apparition has several tens of observations with excessively large residuals. Moreover,  the failure to predict astrometric positions manifests for every possible choice of consecutive apparitions included in the fit.


\subsection{Yarkovsky effect}\label{sec:yarko}
\begin{figure*}[ht!]
\epsscale{0.8}
\plotone{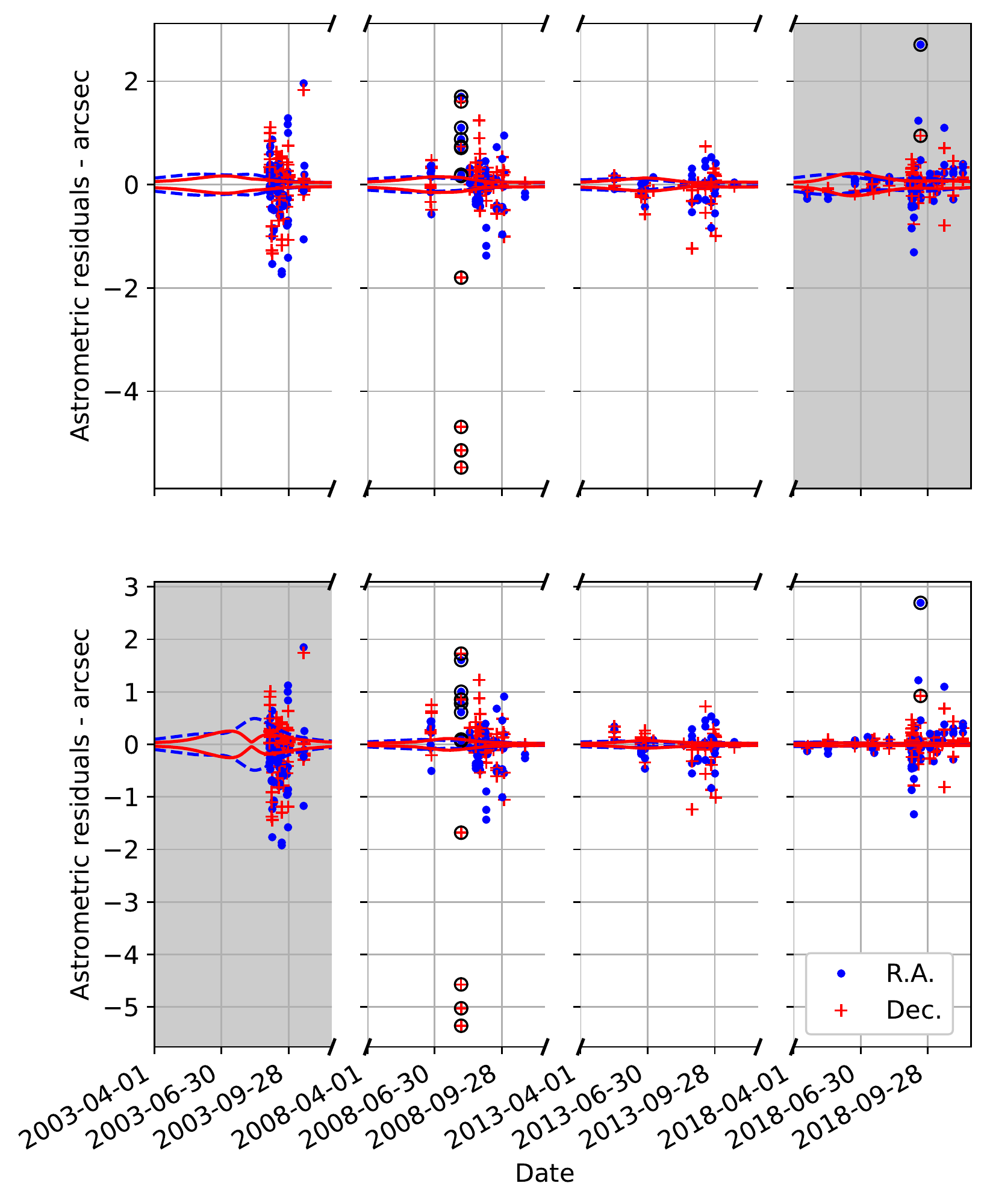}
\caption{Astrometric residuals against the best fit to the astrometry either from 2003 to 2013 (top panel) or from 2008 to 2018 (bottom panel) using a transverse acceleration $A_2/r^2$. Dots correspond to Right Ascension (R.A.) and crosses correspond to Declination (Dec.). Outliers rejected from the fit are indicated with circles. The shaded area corresponds to observations not included in the fit. The dashed line and solid line represent the 1-$\sigma$ ephemeris uncertainties of the orbital solution in R.A. and Dec., respectively. The residuals in the shaded areas are compatible with the ephemeris prediction and thus suggest that the adopted nongravitational model is compatible with the dataset.}\label{fig:ng_nm2}
\end{figure*}

The Yarkovsky effect, a nongravitational perturbation due to anisotropic thermal re-emission of absorbed solar radiation \citep{Vokrouhlicky2015_ast4}, is a reasonable first explanation for the failure to reproduce the data. 
In order to model this effect, we include a simple transverse acceleration $A_2(\text{1 au}/r)^2$, where $r$ is the heliocentric distance and $A_2$ is an estimable parameter \citep{Farnocchia2013_yarko}.
The fit to the full observational dataset is now satisfactory with a $\chi^2$ of 145.1 and a weighted RMS of 0.50. The best fit produces an estimate $A_2 = (212.7 \pm 3.6) \times 10^{-14}$ au/d$^2$, i.e., a detection with signal-to-noise ratio of 58.
This estimate is consistent when using subsets of the data arc over three apparitions:
\begin{itemize}
    \item $A_2 =   (212.5 \pm 9.0) \times 10^{-14}$ au/d$^2$ when fitting the first three apparitions from 2003 to 2013;
    \item $A_2 =   (207.9 \pm 7.9) \times 10^{-14}$ au/d$^2$ when fitting the last three apparitions from 2008 to 2018.
\end{itemize}
Figure~\ref{fig:ng_nm2} demonstrates that fits based on three apparitions that include a transverse acceleration $A_2/r^2$ successfully predict the fourth apparition within the uncertainties. However, the magnitude of the transverse acceleration far exceeds that which could feasibly be produced by the Yarkovsky effect.
\citet{Farnocchia2013_yarko} defined $A_{2,exp}$ as a proxy for the expected value of $|A_2|$ for a given asteroid. This approximation is derived from the $A_2$ value measured for (101955) Bennu, scaled according to the expected size of the target asteroid. Bennu is a good reference because if its extreme obliquity and high Yarkovsky detection signal-to-noise ratio.
For 2003~RM, the absolute magnitude $H = 19.8$ leads to a range in diameter from 300 to 650 m assuming an albedo between 5\% and 25\%.
Therefore, $A_{2,exp} < 8 \times 10^{-14}$ au/d$^2$, which is 27 times smaller than the observed acceleration.
Matching the observed acceleration would require either an unrealistically low bulk density ($\sim$ 0.1 g/cm$^3$) or a size of tens of meters, which requires a nonphysical albedo $>10$.

\subsection{Cometary nongravitational perturbations}
Since the Yarkovsky effect is incompatible with the observed acceleration, we considered the possibility that 2003~RM is a comet and that its motion is affected by perturbations due to outgassing. By modeling the transverse acceleration as $A_2 g(r)$ \citep[with $g(r)$ from][]{Marsden1973}, we obtain a satisfactory fit to the data. This fit produces $\chi^2=138.7$,  a weighted RMS of 0.49, and $A_2 = (332.4 \pm 5.7) \times 10^{-14}$ au/d$^2$. In fact, this fit is  slightly better than the Yarkovsky solution obtained in the previous subsection. 
Table~\ref{tab:orbit} shows the corresponding orbit solution.
The Tisserand parameter is 2.96, which is in the range between 2 and 3 typical of Jupiter-family comets \citep{Duncan2004}.

\begin{deluxetable*}{l|lll}[ht!]
\tablecaption{JPL orbit solution 58 for 2003~RM using the \citet{Marsden1973} $g(r)$ function. The heliocentric orbital elements refer to an osculating epoch of 2017 May 5 TDB and are in the IAU76 ecliptic frame \citep{Seidelmann77}. The error bars correspond to 1-$\sigma$ formal uncertainties.
\label{tab:orbit}}
\tablecolumns{4}
\tablehead{
Parameter & Value & Uncertainty & Units}
\startdata
Eccentricity & $0.6013493571$ & $2.39\times 10^{-8}$ & \\
Perihelion distance & $1.1642435313$ &  $6.98\times 10^{-8}$ & au\\
Time of perihelion TDB & $2018$ July $4.6791744$ & $1.08\times 10^{-5}$ & d\\
Longitude of node & $336.72534377$ &  $9.65\times 10^{-6}$ & $^\circ$\\
Argument of perihelion & $324.5461040$ & $1.20\times 10^{-5}$ & $^\circ$\\
Inclination & $10.86102945$ &  $3.75\times 10^{-6}$ & $^\circ$\\
$A_2$ & $332.42\times 10^{-14}$ & $5.68\times 10^{-14}$ & au/d$^2$\\
\enddata
\end{deluxetable*}

\begin{figure}[ht!]
\epsscale{1.1}
\plotone{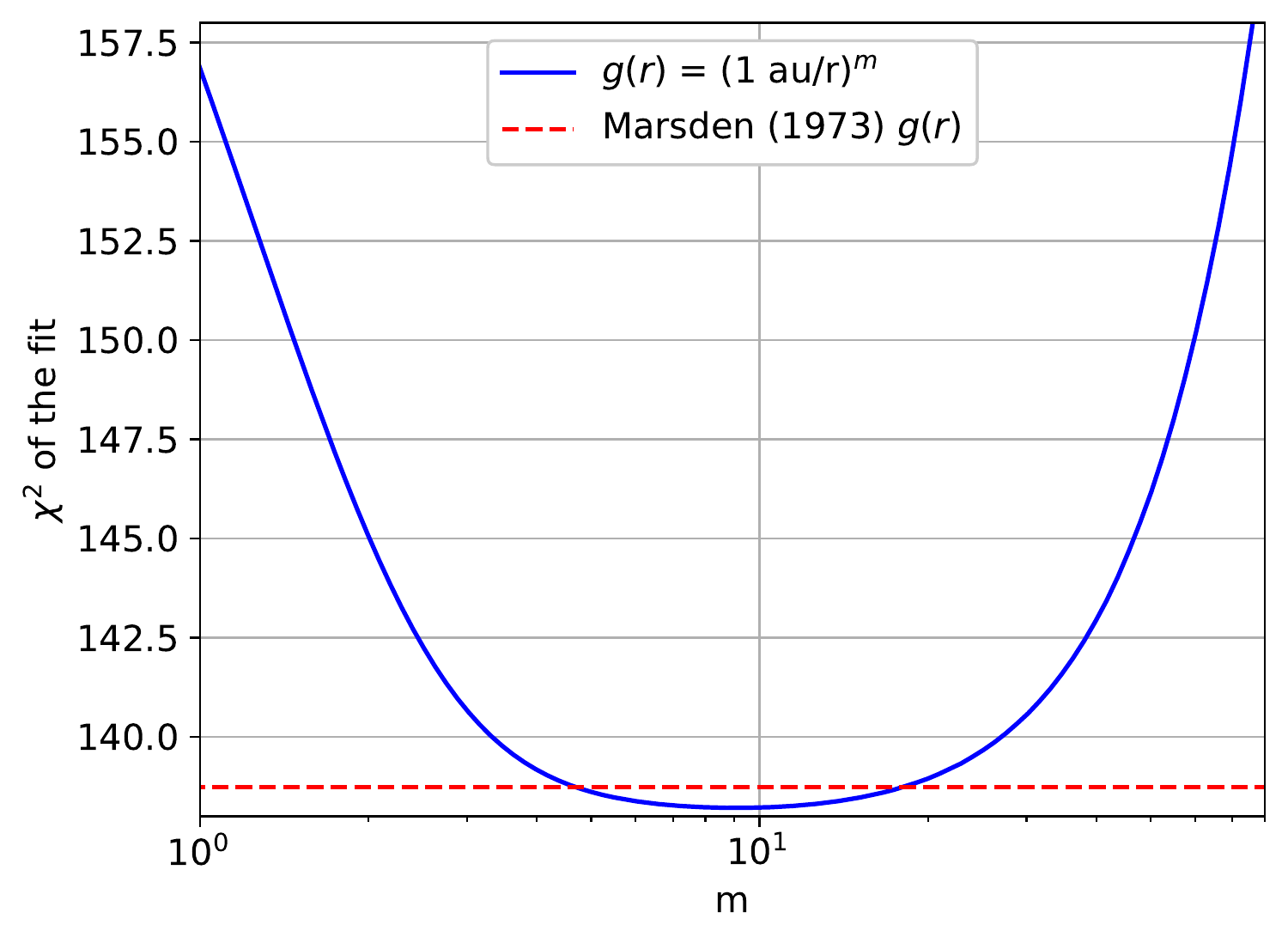}
\caption{$\chi^2$ of the orbital fit as a function of the power law exponent for $g(r) = 1/r^m$.
The full observational arc is included in the fit and $A_2$ is the only nongravitational parameter estimated. The dashed line corresponds to the \citet{Marsden1973} $g(r)$.\label{fig:gr}}
\end{figure}

We also investigate whether or not the fit to the data favors a specific $g(r) \propto 1/r^m$ power law.
Figure~\ref{fig:gr} shows the best-fit $\chi^2$ as a function of $m$.
There is a very shallow minimum for $m=9$ with a broad 3-$\sigma$ range ($\Delta\chi^2 = 9$) of acceptable values from $m=1.6$ to $m=55$.

There is no signal for the radial component $A_1 g(r)$, i.e., using the \citet{Marsden1973} model the estimate of $A_1 = (-2300 \pm   2500)\times 10^{-14}$ au/d$^2$ is compatible with zero within the uncertainty.
The range of possible values for $A_1$ is compatible with $A_1$ being an order of magnitude greater than $A_2$, which is a typical ratio for comets \citep{Farnocchia2014SidingSpring}.

\subsection{Close approach to or collision with another asteroid}
Another possibility is that the orbit of 2003 RM is perturbed by some other small-body perturber.
However, increasing the number of perturbers from 16 to 373 \citep{Farnocchia2021sb441} neither reveals any significant close approaches nor provides a satisfactory fit to any three of the apparitions. While it is possible that some assumed perturber masses are erroneous, we verified that this could not account for the lack of fit.
Specifically, estimating the perturber masses as free parameters within reasonable a priori ranges \citep[similarly to][]{Farnocchia2021Orex} does not allow a fit to the data.
As further proof, the simultaneous estimate of $A_2$ and perturber masses still leads to a 58-$\sigma$ detection of $A_2$. These results clearly favors the additional nongravitational acceleration over corrections to perturber masses.

With an aphelion of 4.7 au, there is a remote possibility that 2003~RM experienced  a close approach or collision with a small body not included as a perturber in the force model.
This close approach or collision would more likely have occurred when 2003~RM crossed the main belt, which is densest on the ecliptic plane \citep[e.g.,][Sec.~6]{Jedicke2015}.
However, the ascending and descending nodes of 2003~RM are at 1.2 au and 3.6 au from the Sun, respectively. The asteroid distribution has low density at these distances.
Still, the 10$^\circ$ inclination of 2003~RM could allow for close encounters or collisions outside of the ecliptic plane.

The $A_2$ signal is present when fitting both the first three apparitions and the last three apparitions.
Therefore, any close approach and collision would need to take place between the second and the third apparition, i.e., between 2008 October 29 and 2013 May 16.
During this time window, we estimated an impulsive velocity variation $\Delta v$ as a function of time as discussed in \citet[][Sec.~5]{Farnocchia2014Nyx}.
To aid the fit to the data, we included the maximum transverse acceleration allowed by the Yarkovsky effect, i.e., $A_2 =
8 \times 10^{-14}$ au/d$^2$ (see. Sec.~\ref{sec:yarko}).

\begin{figure}[ht!]
\epsscale{1.1}
\plotone{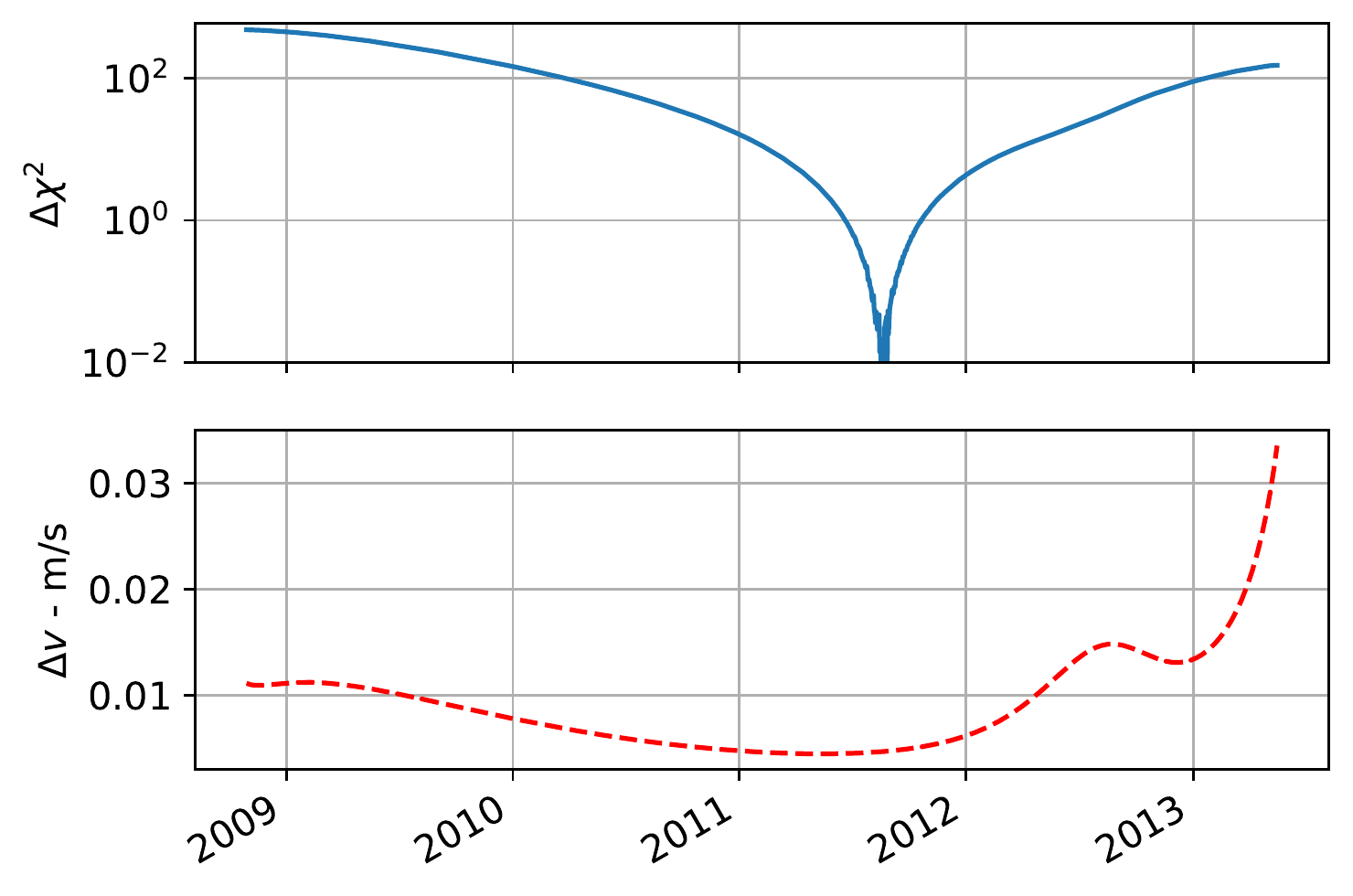}
\caption{Top panel: $\Delta\chi^2$ of the orbital fit for an orbital solution with an impulsive variation in velocity $\Delta v$ as a function of impulse time relative to the lowest $\chi^2 = 285.4$, which corresponds to a $\Delta v$ on 2011 August 19. Bottom panel: magnitude of $\Delta v$ as a function of impulse time.\label{fig:dv_chi2}}
\end{figure}

Fig.~\ref{fig:dv_chi2} shows the $\chi^2$ of the fit as a function of the epoch during which the $\Delta v$ was applied.

There is clear minimum with $\chi^2 = 285.4$ for an impulsive event on 2011 August 19, with a 3-$\sigma$ range from 2011 February 25 to 2012 March 06.

\begin{figure*}[ht!]
\epsscale{0.8}
\plotone{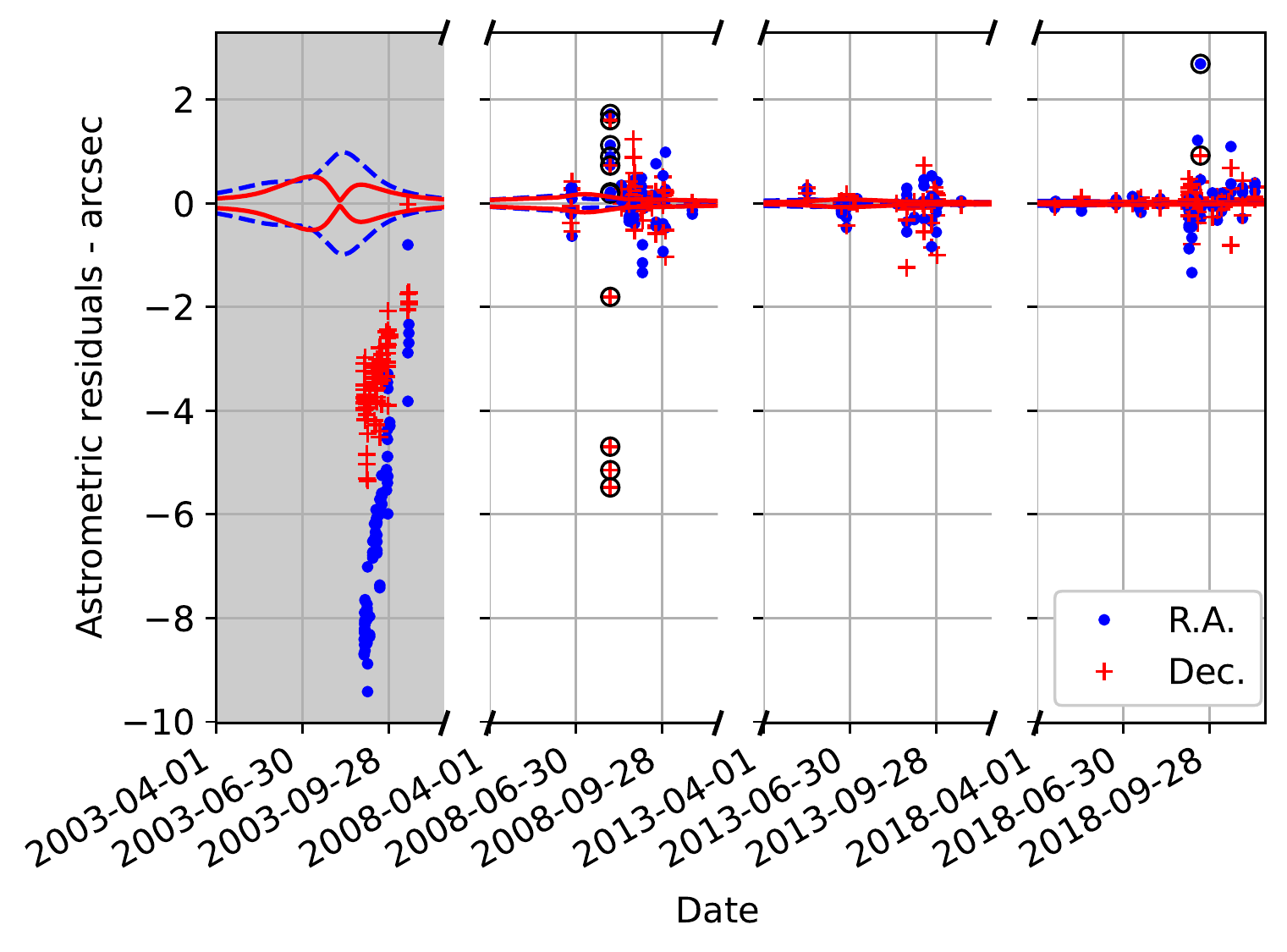}
\caption{Astrometric residuals against the best fit to the astrometry from 2008 to 2018 with an impulsive $\Delta v$ event on 2011 August 9. Dots correspond to Right Ascension (R.A.) and crosses correspond to Declination (Dec.). Outliers rejected from the fit are indicated with circles. The shaded area corresponds to observations not included in the fit. The dashed line and solid line represent the 1-$\sigma$ ephemeris uncertainties of the orbital solution in R.A. and Dec., respectively. The large residuals in the shaded areas far exceed the ephemeris prediction uncertainties and thus imply problems in the force model.}\label{fig:dv_24}
\end{figure*}

We rule out that possibility that an impulsive $\Delta v$ would explain the deviation from a gravity-only model for the following reasons:
\begin{itemize}
    \item During the time frame surrounding the minimum of $\chi^2$, 2003~RM travels from 4.7 au to 4.1 au from the Sun and from 0.5 au to 0.1 au above the ecliptic plane. Because this region of the Solar System has a low density of asteroids \citep{Lagerkvist1997,Jedicke2015}, it is extremely unlikely that an impulsive $\Delta v$ event occured there.
    \item The $\chi^2$ of the fit is significantly higher ($\Delta\chi^2 \sim 147$) than that obtained with the transverse nongravitational acceleration model. This is true despite the fact that a larger number of parameters were used: four (epoch and components of $\Delta v$) instead of one ($A_2$).
    \item A three-apparition orbit solution with $\Delta v$ estimated provides poor predictions for the fourth apparition.
    For example, Fig.~\ref{fig:dv_24} shows the astrometric residuals against a solution based on the last three apparition that estimates a $\Delta v$ on 2011 August 19. The residuals of the 2003 apparitions are much larger than the ephemeris prediction uncertainties.
\end{itemize}

\section{Search for evidence of cometary activity} \label{sec:cometary}
No observer has reported a clear detection of cometary activity for 2003~RM to the Minor Planet Center. As a result 2003~RM is classified as an asteroid. 
We carefully reviewed our observations (see Sec.~\ref{sec:astrometry}) and 2003~RM appears stellar in all of them, thus providing no observational evidence of a cometary nature for this object. Below we provide a detailed analysis for two of the observations with the highest signal-to-noise ratio (SNR): Pan-STARRS observations on 2013 August 17, 34 days past perihelion, when 2003~RM was at heliocentric distance of 1.2 au, and VLT observations on 2018 September 19, 76 days past perihelion, when 2003~RM was at a heliocentric distance 1.5 au.






\subsection{Pan-STARRS} 

\begin{figure}
    \centering
    \includegraphics[width=8.5cm]{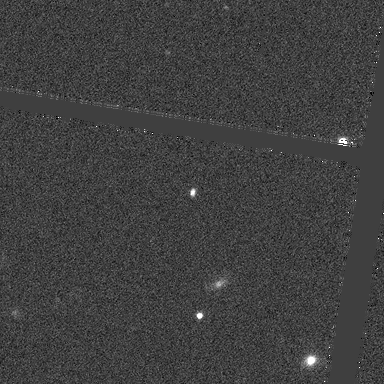}
    \caption{2003~RM (in the center, North is up, East is left) as it appeared on 2013 August 17 from Pan-STARRS1 moving North at $0.62$ degrees per day towards a cell gap.  Comparison of the FWHM with visible stars in a larger region than shown here indicates it was not active. At the observation time, 2003 RM was at 1.2 au from the Sun and 0.3 au from Earth.}
    \label{fig:activity}
\end{figure}

We searched the image archive of the $1.8$~m Pan-STARRS1 telescope \citep{Wainscoat2020,Chambers2016} located on Maui, Hawai'i. We inspected the ``chip'' images of all matching exposures for low-level cometary activity. These images are normally ``warped'' to a common fixed plane-of-sky projection with $0.25''$ pixels (to allow for deep stacks for non-asteroid science) for survey operations, but using them allows for inspection of the native $0.256''$ pixel data without resampling, which can be useful in identifying sub-pixel coma. 

We employed the \citet{Veres2012} algorithm to fit the point-spread-function (PSF) of both 2003~RM and all visible field stars that appear in the Gaia-DR2 \citep{Gaia18} catalog. The full width at half maximum (FWHM) was the primary metric to search for activity. The fit primarily uses an asymmetric Gaussian function and adopts a trailed Gaussian for fast moving  objects. This method also reports uncertainties in the fit. For trailed PSFs, the FWHM corresponds to the direction perpendicular to the motion. The ``curve of growth'' aperture flux (essentially a plot of aperture flux as a function of aperture radius) is dependent on the PSF of a given object and, for large-enough radii, allows for the identification of faint coma when compared to comparably bright field stars which do not show such coma. 

\begin{figure*}[ht!]
\centering
\includegraphics[width=14.0cm]{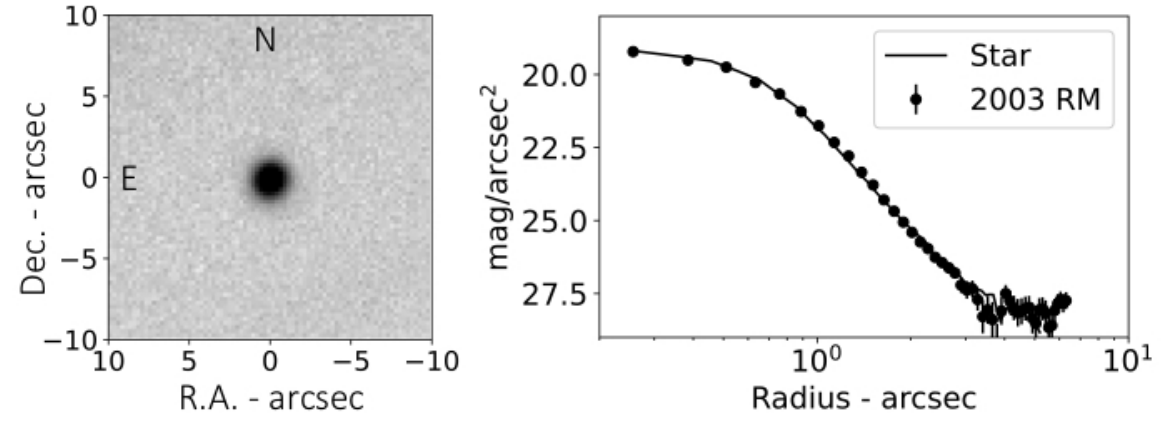}
\caption{Left: Stack of 82 VLT frames totalling 3280~s exposure time on 2003~RM.
The observations were collected on 2018 September 19, when 2003 RM was at 1.2 au from the Sun and 0.3 au from Earth.
The image is 20$''$ wide. The individual frames were subtracted for background objects. Right: Surface brightness profile of 2003~RM (dots) and of a scaled field star (line). \label{fig:VLTprof}}
\end{figure*}

Unfortunately, there do not exist many Pan-STARRS images for 2003~RM. However, the images with higher SNR show it to be consistent with a star-like appearance.  
On 2013 August 17, a $60$ second z-band image with SNR $\sim 15$ had a FWHM of $1.01'' \pm 0.03''$ measured perpendicular to the direction of motion. Since the plane-of-sky motion of 2003~RM was aligned to the stellar minor axis (ie: there is a slight asymmetry in the PSFs), the major axis FWHM of $0.99'' \pm 0.06''$ must be used for comparison.  The object was too trailed for a curve-of-growth comparison, and is illustrated in Figure \ref{fig:activity}.
On 2013 September 14, a $43$-second g-band image with SNR $\sim 8$ had a FWHM of $1.22'' \pm 0.05''$, with the motion of 2003~RM aligned with the stellar major axis.  The stellar FWHM minor axis for comparison was $1.24'' \pm 0.11''$. 
On 2013 October 24, a pair of $45$-second i-band images with SNR $\sim 6$ were stacked to have a FWHM of $1.01'' \pm 0.05''$ (minor axis) and $1.09'' \pm 0.06''$ (major axis) compared to stellar FWHM $1.00'' \pm 0.06''$ and $1.15'' \pm 0.04''$ respectively.  The curve of growth in each image was comparable, although the object was fainter than the field stars.
On 2018 October 17, a $45$-second w-band image with SNR $\sim 8$ had a FWHM of $1.38'' \pm 1.06''$ (minor axis) and $1.58'' \pm 0.07''$ (major axis) evidently consistent with the stellar FWHM of $1.33'' \pm 0.06''$ and $1.49'' \pm 0.06''$ respectively.  The curve-of-growth was consistent with field stars of similar brightness.

Based on these images, there is no evidence of a non-stellar appearance, and an object like this would never be reported even as a marginal comet candidate.

\subsection{Very Large Telescope}

On 2018 September 19, 2003~RM was observed with the Unit Telescope~1 of the ESO Very Large Telescope (VLT) on Paranal, Chile, with the FOcal Reducer and low dispersion Spectrograph~2 (FORS2) \citep{fors94}. These observations did not use a filter in order to reach the deepest possible magnitude and surface brightness for an object with solar color. 
A total of 82 randomly dithered exposures of 40~s were acquired in Service Mode on 2018 September 19. They were bias and flat corrected (with a twilight flat) and normalized to an exposure time of 1~s.  Remaining low spatial frequency residuals were removed by dividing the frames again by a ``super flat field.'' This was obtained using a median of the science frames -- after normalization of the sky level around the expected position of the object and masking the background objects.

These frames were then co-aligned using a dozen field objects as reference. A master ``star'' stack was produced, which was used to calibrate the field astrometrically using stars from the Gaia DR2 catalogue \citep{Gaia16, Gaia18}. The expected pixel position of 2003~RM on each frame was computed. The master star stack was then subtracted from each of the individual frames. This subtraction produced images in which the majority of the contribution from background objects was removed. The original and star-subtracted frames were then stacked after being re-centered on the expected position of the object, using an average with outlier rejection. Another stack was produced, including only the 55 exposures where the expected position of the object was at least $10''$ from a background object. 

Both the total stack of star-subtracted frames (displayed in Fig.~\ref{fig:VLTprof}) and the partial ones (totalling 3280~s and 2200~s exposure time, respectively) were searched for dust. Visual inspection of the frame reveal no visible extension of the object.


A photometric profile of 2003~RM was produced by integrating its light in a series of circular apertures centered on the object with increasing radii. The instrumental fluxes in these apertures were converted to surface brightness in the concentric rings. These were then converted to magnitudes using a photometric zero point of 27.8 \citep[from][for filter-less observations]{Hainaut+21}. The profile of a field star slightly brighter than 2003~RM was obtained in a similar manner from the ``star'' stack, and scaled to the brightness of the object for comparison. These profiles are presented in Fig.~\ref{fig:VLTprof}, showing no divergence down to 28~mag/sq.~arcsec.

To quantify the lack of dust, it is worth noting that the profile of the object is perfectly stellar within the noise level. Dust contributing to up to 5-$\sigma$ could exist within the error bars. Integrating the corresponding flux from 0.5$''$ to $2''$ corresponds to a total magnitude of 24.0. Assuming dust grains with a radius of $1\mu$m and with a density of 3000~kg$/$m$^{3}$ and an albedo of 0.2, we derive a coma mass limit below 4~kg. This limit should be taken as an order of magnitude estimate based on the numerous assumptions.

\section{Heliocentric Light Curve Modeling} 

Evidence of cometary activity in the form of scattered light from the dust and nucleus can be found from the shape of the heliocentric light curve even in the absence of imaged dust. This can occur when there is brightening caused by light scattered from dust contained within the seeing disk, as in the case of a gravitationally bound dust coma such as the inner coma of (2060) Chiron \citep{Meech1990} and perhaps (468861) 2013 LU28 \citep{Slemp2022} and (418993) 2009 MS9 \citep{Bufanda2022}.

To derive limits on outgassing, we used the ice sublimation model to compare the heliocentric light curve brightness to a predicted comet dust production from outgassing. The model computes the amount of gas sublimating from an icy surface exposed to solar heating, as described in detail in \citet{Meech2017}. The total brightness within a fixed aperture combines radiation scattered from the nucleus and dust entrained in the sublimating gas flow and dragged from the nucleus. The model free parameters include the ice type, nucleus radius, albedo, emissivity, density, dust properties (sizes, density, phase function), and fractional active area. 2003 RM has not been well-characterized so we have significant uncertainties for all of these parameters. 

\begin{figure}[ht!]
\centerline{
\includegraphics[width=8.8cm]{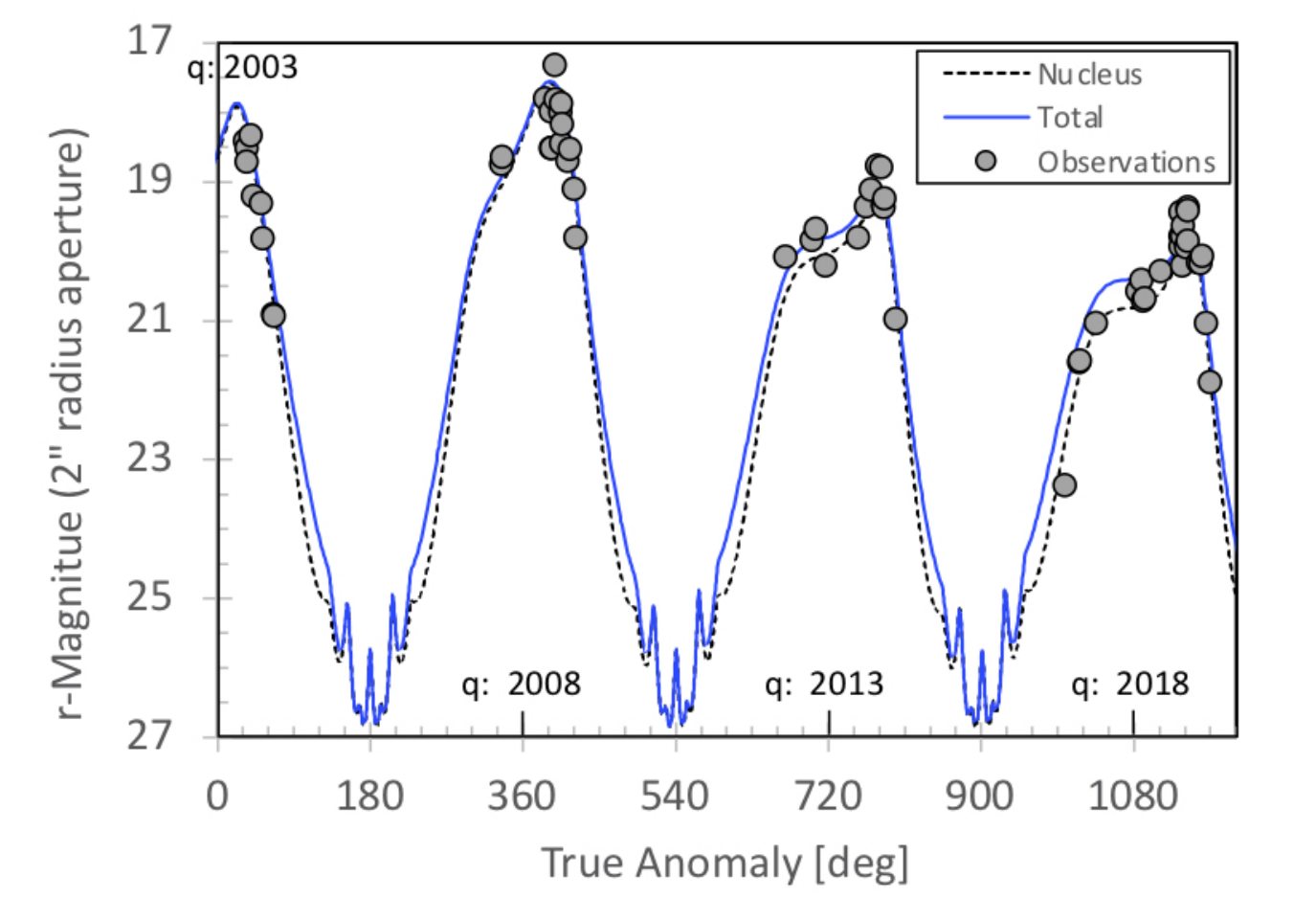}
}
\caption{The best-fitting H$_2$O-sublimation model allows for brightening due to activity consistent with the scatter in the data. The photometric data cover the four observing apparitions from 2003 to 2018 described in Sec.~\ref{sec:astrometry}.}
\label{fig:model}
\end{figure}

We used magnitudes from 329 observations as reported to the Minor Planet Center between September 2003 and November 2018 to model the heliocentric light curve. This data represents observations from 34 different observatories reported using several different photometric systems, including some sets with no filter listed.  In order convert all the measurements to a Sloan (SDSS) r-band, we needed to know surface color (e.g. the taxonomic class) for 2003 RM. Most NEOs belong to the S taxonomic class \citep{Ieva2020}, and based on this assumption we used the mean SDSS colors from \citet{Ye2016} and the transformations from \citet{Jordi2010} and Lupton (2005)
\footnote{\url{https://classic.sdss.org/dr4/algorithms/sdssUBVRITransform.html\#Lupton2005}} to convert the G, V, and R filters to the SDSS r-band.  Because there was significant scatter in the data (which was likely never intended to represent high-precision photometry), we averaged data points taken on the same night. Data from five of the observing stations were significantly discrepant with observations taken at the same time and were not used. The resulting photometry is shown in Fig.~\ref{fig:model} and has a an error of $\sim$0.3 mag (about the size of the data points).

In Sec.~\ref{sec:past} we show that 2003~RM may be associated with the Eos family which maps to the K-taxonomic class, a subset of the S-type members. These asteroids tend to have the same visible colors as S-types but slightly lower albedos, around $p_v$ $\sim$ 0.12 \citep{Clark2009}. Table~\ref{tab:model_parms} lists the starting parameters for the model, along with the references for the starting values.  Some of these are based on estimates for S-type asteroids and others are relevant to active comets.

\begin{deluxetable*}{l|c|cc|ccc}[bht!]
\scriptsize
\tablewidth{0pt}
\tablecaption{Summary of Sublimation Model Parameters \label{tab:model_parms}}
\tablecolumns{6}
\tablehead{
\multicolumn{2}{c}{ } & \multicolumn{2}{|c|}{Initial} & \multicolumn{3}{|c}{Fit}}
\startdata
\multicolumn{2}{l|}{Sublimation Parameter}        & Value & Source$^{\dag}$ & H$_2$O & CO$_2$-Limit & CO-Limit  \\ \hline
Nucleus Radius [km]              & R$_N$          & 0.150 & [1]  & 210   & 210  & 210  \\ 
Emissivity                       & $\epsilon$     & 0.95  & [2]  & 0.9   & 0.9  & 0.9  \\
Nucleus Phase function [mag/deg] & $\beta_{nuc}$  & 0.04  & [3]  & 0.05  & 0.05 & 0.05 \\ 
Nucleus Density [kg/m$^{3}$]     & $\rho_n$       & 1900  & [4]  & 1900  & 1900 & 1900 \\ 
Nucleus Albedo                   & p$_v$          & 0.25  & [1]  & 0.12  & 0.12 & 0.12 \\ 
Coma Phase function [mag/deg]    & $\beta_{coma}$ & 0.02  & [5]  & 0.02  & 0.02 & 0.02 \\ 
Grain Density, [kg/m$^{3}$]      & $\rho_g$       & 3000  & [6]  & 3000  & 3000 & 3000 \\
Grain Radius, [$\mu$m]           & a              &    1  & [7]  &    1  &    1 &    1 \\ 
Fractional Active Area           & FAA            & 0.04  & [7]  & 0.003 & 0.00006 & 0.0002 \\
\hline
Inferred gas production [molec/s]& Q              &       &      & 3.8E24& 1.7E23 & 7.3E23 \\
\enddata
\tablecomments{$^{\dag}$[1] matches observed H-value for high albedo S-types; [2] measured for 67P by Rosetta \citep{Spohn2015}; [3] measured for an S-type \citet{Gehrels1977}; [4] average density for small S-type asteroids ((25143) Itokawa, (101955) Bennu, (162173) Ryugu, (65803) Didymos); [5] measured for comets \citep{Meech1987}; [6] see discussion in \citet{Meech1997}; [7] Assumed starting values for comets.}
\end{deluxetable*}

\vspace{-0.8cm}
A good fit to the data is shown in Fig.~\ref{fig:model}, and the fit parameters are shown in Table~\ref{tab:model_parms} for water. Because of all the nucleus uncertainties, we varied R$_N$, $\beta_{nuc}$, and FAA. 
The total curve is the brightness combined from the nucleus and the dust lifted off by the water sublimation model. In Fig.~\ref{fig:model}, the nucleus brightness is the best fit to the data and the ``total'' model includes scattered light from the nucleus and the dust coma. The model is shown plotted against true anomaly (TA=0$^{\circ}$ is at perihelion). Because the data span four apparitions, TA increases by 360$^{\circ}$ at each perihelion. The VLT data were taken about 2.5 months post-perihelion, at TA=1144$^{\circ}$. The total model curve allowing for activity is consistent with the envelope of the scatter in the data. The implied dust production rate at perihelion for the fourth apparition from this model is 0.114 kg/s and converting this to a gas production rate assuming a dust-to-gas ratio of 1 \citep{Marschall2020} gives an inferred Q(H$_2$O) $\sim$ 3.8$\times$10$^{24}$ molec/s.\footnote{We have no knowledge of the dust-to-gas ratio. If the dust-to-gas ratio were larger then the model would produce more dust per unit of gas production.  The light curve in Fig.~\ref{fig:model} is a measurement of scattered light from the nucleus and dust.  A larger dust-to-gas ratio would imply lower estimates for the gas.} If the photometric scatter were much smaller (e.g. less than 0.01 mag), then the model would be sensitive to activity at the level of $\sim$ 10$^{23}$ molec/s. At this gas production level the integrated brightness from 1 $\mu$m dust contained within a 2$''$ radius aperture would be $r$-mag 24, consistent with the limits from the VLT data. This limit is just above the amount of gas production that is required for a nucleus of this size and density based on the nongravitational accelerations (see Fig.~\ref{Fig:h2o} below). 

In order to assess how much of the scatter in the data might be due to the object's rotation, we examined the time series individual images taken over 2 hours from the VLT on 2018 September 19. We used the flattened sky-subtracted frames and excluded the frames where 2003 RM was too close to field stars (during the first hour). The remaining data shows a brightness variation of $\sim$0.05 mag with about 14 cycles over a span of 50 minutes that, if statistically significant, would suggest a rapid rotation period of 2003 RM. More data would be necessary to investigate this further. 

These models were run on the assumption that a body this small may no longer contain ices as volatile as CO$_2$ or CO, and the gas production rate limits, which are lower than that for water, are also shown in Table~\ref{tab:model_parms}.  However, as discussed in Sec.~\ref{sec:past}, the dynamical lifetime of NEOs on orbits like that of 2003 RM is relatively short. Comet 2P/Encke has an even smaller perihelion than 2003 RM ($q$ = 0.33 au) with a slightly larger aphelion ($Q$ = 4.09 au). Comet Encke could have migrated in and become decoupled from Jupiter on a similar timescale. Comet Encke exhibits a curious behavior: near perihelion it often has very little visible dust, but significant gas and is often active at aphelion \citep{meech2001, Fink2009}. However, this comet still has a very strong CO$_2$ or CO production \citep{reach2013}. For both CO$_2$ and CO, similar limiting production rates of $<$10$^{24}$ molec/s are found based on the heliocentric light curve.


\section{Required Outgassing Production}\label{sec:constraints}



We consider a variety of assumptions regarding the nature of the outgassing. The production  depends on several unconstrained factors, including the albedo (and corresponding size) of the body, the temperature of the outgassing gas, the extent to which the outflowing gas is collimated vs. isotropic,  and the dust-to-gass mass ratio.

The production rate $Q(X)$ of a given species, $X$, can be calculated via the conservation of linear momentum --- essentially just the ``rocket'' equation --- using the following equation: 

\begin{equation}\label{eq:production_total}
    Q(X) = \,\bigg(\,\frac{M_{\rm Tot} }{m_{\rm X}}\,\bigg)\,\,\bigg(\,\frac{ |\ddot{\mathbf r}|}{v_{\rm Gas} \zeta}\,\bigg)\,.
    \end{equation}
In Equation \ref{eq:production_total}, the total mass is given by $M_{\rm Tot} = \rho_{\rm Bulk} V$, where, $\rho_{\rm Bulk}$  and  $V$  are the bulk density and  the volume of the nucleus. $\ddot{\mathbf r}$ is the instantaneous acceleration at the heliocentric position. $m_{\rm X}$ is the mass of the outgassing molecule and $v_{\rm Gas}$ is the gas outflow velocity. $v_{\rm Gas}$ can be related to the temperature of the outflowing species by equating the kinetic energy to the thermal energy, which is calculated using:

\begin{equation}\label{eq:vgas}
    v_{\rm Gas} = \,\bigg(\, \frac{8k_B T_\mathrm{Gas}}{\pi m_\mathrm{X}}\,\bigg)^{1/2}\,.
\end{equation}
In Equation \ref{eq:vgas}, $T_\mathrm{Gas}$ is the temperature of the outgassing species. In Equation \ref{eq:production_total}, the factor $\zeta$ parameterizes the geometry of the outgassing species.  $\zeta$ = 1 corresponds to a fully collimated outflow and $\zeta$ = 0.5 corresponds to an entirely isotropic hemispherical outflow. For a complete definition of the parametric $\zeta$ function, see Section 4.1 in \citet{Jewitt2022}.

\begin{figure}
\begin{center}
\includegraphics[scale=0.4,angle=0]{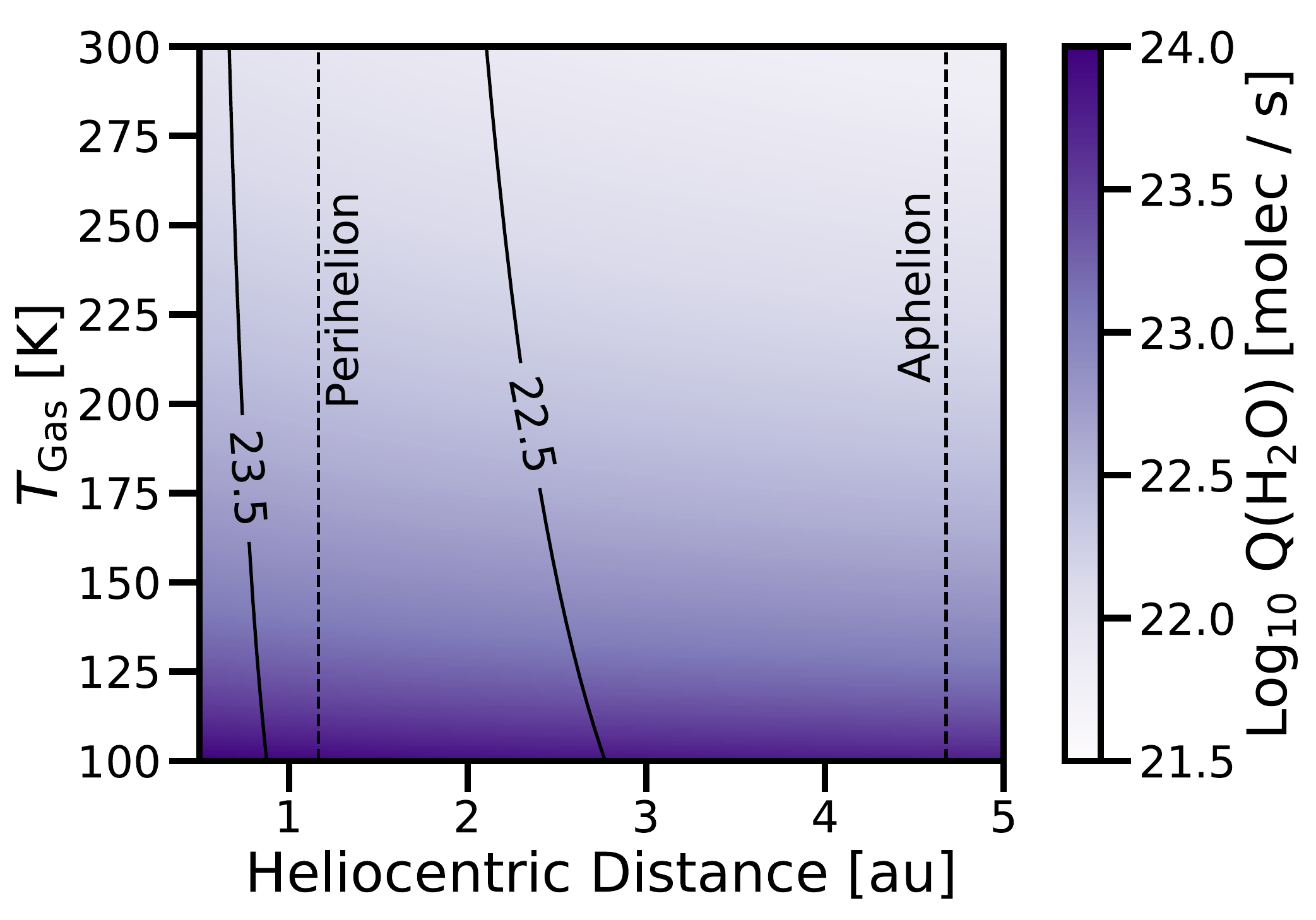}
\caption{The H$_2$O production rates required to produce the measured nongravitational acceleration of 2003 RM. We calculate this value using Equations \ref{eq:production_total} and \ref{eq:vgas},  with  $\zeta=1$, radius of $210$ meters, $\rho_{\rm Bulk} = 1.9$ g cm$^{-3}$ and   $A_2 = 213.7 \times 10^{-14}$ au/d$^2$ for a $(r/1\text{ au})^{-2}$ dependency for heliocentric distance $r$.}\label{Fig:h2o}
\end{center}
\end{figure}

With this construction, we calculate the production rate required to produce the measured nongravitational acceleration under the assumption of a pure H$_2$O outgassing. We show the resulting production rates for a range of outgassing temperature and heliocentric distance in Figure \ref{Fig:h2o}. In order to calculate this production, we assume that the outflow is completely collimated and $\zeta=1$, a spherical nucleus with radius of $R_{\rm Nuc}=210$ m, a bulk density of $\rho_{\rm Bulk} = 1.9\ {\rm g\, cm^{-3}}$ and   the best fit nongravitational acceleration  $A_2 = 213.7  \times 10^{-14}$ \text{ au}/d$^2$ for a $ (r/1\text{ au})^{-2}$ dependency on heliocentric distance.

This production rate calculation can easily be generalized to other assumptions of the outflow and its composition. The following scaled equation:

\begin{equation}\label{eq:production_total_scaled}\begin{split}
    Q(X) \simeq 2.4\times10^{23} {\rm \, molec\,s^{-1}}    \,\bigg(\,\frac{\rho_{\rm Bulk} }{1.9 {\rm \,g\,cm^{-3}}}\,\bigg)\,\\ \,\bigg(\,\frac{R_{\rm Nuc} }{210 {\rm m}}\,\bigg)^3\, 
    \,\bigg(\,\frac{m_{\rm H_2O}}{m_{\rm X}}\,\bigg)^{1/2}\, \,\bigg(\, \frac{1}{\zeta}\,\bigg)\, \,\bigg(\, \frac{100 \,{\rm K}}{T_{\rm Gas}}\,\bigg)^{1/2}\, \\ \,\bigg( \frac{|A_2|}{213.7  \times 10^{-14}  {\rm \,au\,d^{-2}}} \,\bigg)\, \,\bigg(\,\frac{1 {\rm \, au}}{r}\,\bigg)^2\,
    \end{split}
    \end{equation}
may be used to estimate production rates for a variety of assumptions.

\section{Dynamical history} \label{sec:past}

2003~RM is currently located adjacent to the Eos family in terms of its semimajor axis and inclination. Therefore, the most likely source region for the object is in the mid-to-outer part of the main asteroid belt or among Jupiter-family comets.  We used the medium resolution version of the evolutionary Near-Earth Object (NEO) population model by \citet{Granvik2018} to quantify our assessment of the source region. The NEO population model allows us to estimate the relative likelihood that 2003~RM would have entered the NEO region through one of the seven escape regions described by that model. The input for the model assessment are 2003~RM's orbital elements ($a$, $e$, $i$) and absolute magnitude ($H$). According to this model, 2003~RM has an $(86 \pm 4)$\% and $(11 \pm 4)$\% probability of originating in the mid-to-outer asteroid belt and Jupiter-family comet region, respectively. An origin in the inner part of the asteroid belt, including the Hungaria or Phocaea groups, is less likely, with a probability of less than 4\% in total. Thus, it appears that an asteroidal origin is more likely than a cometary origin for 2003~RM. It is worth pointing out that a main belt origin is not necessarily incompatible with a cometary nature. In fact, \citet{Hsieh2020} show how Jupiter-family comets can originate in the main belt.

We used the model developed by \citet{Toliou2021} to assess 2003~RM's orbital history and lower perihelion distance, $q$. Based on the orbital elements and absolute magnitude, the model predicts that 2003~RM attained $q<1\,$au and $q<0.5\,$au with probability of $(42 \pm 2)$\% and $(5 \pm 2)$\%, respectively. Therefore, it is likely that 2003~RM has never experienced heating beyond that which it would have been exposed to at about 0.5~au.


How much solar heating and volatile release are likely to have occurred in 2003~RM's past?
The average lifetime of NEOs originating in the mid-to-outer asteroid belt is 300--400~kyr. Approximately 80\% of these NEOs will eventually be ejected from the inner Solar System as a result of a close encounter with Jupiter \citep{Granvik2018}. The median duration of the time that an object---currently with a 2003~RM-like orbit--- spends on orbits with $q<0.5\,$ au, $q<1\,$ au, and $q<1.3\,$ au is about 40, 100, and 190~kyr, respectively \citep{Toliou2021}. Because only a small fraction of the total orbital period occurs close to perihelion, these median times should be reduced by orders of magnitude to estimate the time spent at heliocentric distances $r<0.5\,$au, $r<1\,$ au, and $r<1.3\,$ au, respectively. If 2003~RM initially contained volatiles, it is possible that the object retained some fraction of them while being an NEA based on these timescales. Moreover, it is believed that main-belt comets that are typically found in the mid-to-outer asteroid belt initially contained volatiles. Therefore, the excess nongravitational acceleration measured for 2003~RM could be caused by weak activity. This is similar to what might have occurred on 1I/`Oumuamua.


\section{Conclusions} \label{sec:discussion}
Based on an observation arc from 2003 to 2018, it is clear that the motion of near-Earth asteroid 2003 RM is affected by significant nongravitational perturbations in the transverse direction.
Although transverse nongravitational accelerations on asteroids are typically caused by the Yarkovsky effect, the magnitude of the required anomalous acceleration is much greater than can be explained by this phenomenon. We investigated and ruled out alternative explanations for the observed orbital deviations such as close approaches or a collision with another asteroid.
Therefore, we conclude that the most likely source of this nongravitational perturbation seems to be some form of cometary outgassing, orders of magnitude smaller than what is typically observed in comets, and that 2003 RM could be a cometary object.  (See Extended Data Fig.~1 from \citeauthor{Micheli2018}\ \citeyear{Micheli2018} and Fig.~4 from  \citeauthor{Farnocchia2014SidingSpring}\ \citeyear{Farnocchia2014SidingSpring}.)


However, direct imaging does not reveal any indication of cometary activity and even a statistical analysis of the dynamical history of 2003 RM favors an asteroidal origin.
If 2003 RM is actively sublimating, it is curious that the object does not display a bright cometary tail. We performed a detailed search through observational data, especially from Pan-STARRs and the VLT, and we found no evidence for extended coma or brightening events in the secular light curve. Therefore, if this object is sublimating, the dust coma is very weak or non-existent.

We estimated the required levels of H$_2$O outgassing that would be required to produce the  nongravitational perturbations. By invoking the conservation of linear momentum, we estimated that 2003 RM requires production rates of order $Q(\mathrm{H}_2\mathrm{O})\sim 10^{22}-10^{23}$ molec/s. We also compared the photometric lightcurve of 2003 RM with a model of (i) a bare nucleus and (ii) a nucleus with the addition of scattered light from a dust coma contained within the unresolved seeing disk. We found that the scatter in the data could be explained by an inferred outgassing rate of Q(H$_2$O)$<$ 10$^{24}$ molec/s, a limit just above the amount of gas production required to produce the nongravitational perturbations. Therefore, it is possible but not definitive that 2003 RM could be exhibiting a low level of activity. As a point of reference, the upper limit for main belt comet water outgassing near perihelion is between 10$^{24}$ and 10$^{26}$ molec/s \citep{Snodgrass2017}.


With the information currently available, the nature of 2003 RM remains a puzzle and it is not clear if the object should be considered as an asteroid or a comet, if not as a new class of small body in the Solar System. 2003 RM is not alone in this regard. In an accompanying paper (Seligman et al. submitted), we report the discovery of similar nongravitational accelerations in six other small objects in the Solar System. Even more strikingly, the first discovered interstellar object 1I/`Oumuamua also had a significant nongravitational acceleration but no coma.

Additional observations of 2003 RM are necessary to understand the origin of the detected large nongravitational perturbations, which would in turn help solve the similar puzzle on `Oumuamua's nature. Specifically, high signal-to-noise ratio space-based near- and mid-infrared observations could be sensitive to detection H$_2$O, CO$_2$ and CO outgassing activity even in the absence of micron sized dust particles. Observations with the \textit{James Webb Space Telescope} would be particulary helpful for identifying the source of the acceleration of 2003 RM. 


\section{Acknowledgments}
Part of this work was carried out at the Jet Propulsion Laboratory, California Institute of Technology, under a contract with the National Aeronautics and Space Administration (80NM0018D0004). DZS acknowledges financial support from the National Science Foundation  Grant No. AST-17152, NASA Grant No. 80NSSC19K0444 and NASA Contract  NNX17AL71A from the NASA Goddard Spaceflight Center. KJM, JVK and JTK acknowledge support through and award from the National Aeronautics and Space Administration (NASA) under grant NASA-80NSSC18K0853.
This research has made use of data and/or services provided by the International Astronomical Union's Minor Planet Center. This research has made use of observations collected at the European Southern Observatory under ESO programme 2101.C-5049(A).
 Pan-STARRS is supported by the National Aeronautics and Space Administration under Grant No. 80NSSC18K0971 issued through the SSO Near Earth Object Observations Program.

Copyright 2022. California Institute of Technology.


\bibliography{2003rm}{}

\begin{thebibliography}{}
\expandafter\ifx\csname natexlab\endcsname\relax\def\natexlab#1{#1}\fi
\providecommand{\url}[1]{\href{#1}{#1}}
\providecommand{\dodoi}[1]{doi:~\href{http://doi.org/#1}{\nolinkurl{#1}}}
\providecommand{\doeprint}[1]{\href{http://ascl.net/#1}{\nolinkurl{http://ascl.net/#1}}}
\providecommand{\doarXiv}[1]{\href{https://arxiv.org/abs/#1}{\nolinkurl{https://arxiv.org/abs/#1}}}

\bibitem[{{Appenzeller} {et~al.}(1998){Appenzeller}, {Fricke}, {F{\"u}rtig},
  {G{\"a}ssler}, {H{\"a}fner}, {Harke}, {Hess}, {Hummel}, {J{\"u}rgens},
  {Kudritzki}, {Mantel}, {Meisl}, {Muschielok}, {Nicklas}, {Rupprecht},
  {Seifert}, {Stahl}, {Szeifert}, \& {Tarantik}}]{fors94}
{Appenzeller}, I., {Fricke}, K., {F{\"u}rtig}, W., {et~al.} 1998, The
  Messenger, 94, 1

\bibitem[{{Bufanda} {et~al.}(2022){Bufanda}, {Meech}, {Kleyna}, {Hainaut},
  {Bauer}, {Stephens}, {Veres}, {Micheli}, {Keane}, {Weryk}, {Wainscoat},
  {Sahu}, \& {Bhatt}}]{Bufanda2022}
{Bufanda}, E., {Meech}, K.~J., {Kleyna}, J.~T., {et~al.} 2022, arXiv e-prints,
  arXiv:2211.02664.
\newblock \doarXiv{2211.02664}

\bibitem[{{Carpino} {et~al.}(2003){Carpino}, {Milani}, \&
  {Chesley}}]{Carpino2003}
{Carpino}, M., {Milani}, A., \& {Chesley}, S.~R. 2003, \icarus, 166, 248,
  \dodoi{10.1016/S0019-1035(03)00051-4}

\bibitem[{{Chambers} {et~al.}(2016){Chambers}, {Magnier}, {Metcalfe},
  {Flewelling}, {Huber}, {Waters}, {Denneau}, {Draper}, {Farrow}, {Finkbeiner},
  {Holmberg}, {Koppenhoefer}, {Price}, {Rest}, {Saglia}, {Schlafly}, {Smartt},
  {Sweeney}, {Wainscoat}, {Burgett}, {Chastel}, {Grav}, {Heasley}, {Hodapp},
  {Jedicke}, {Kaiser}, {Kudritzki}, {Luppino}, {Lupton}, {Monet}, {Morgan},
  {Onaka}, {Shiao}, {Stubbs}, {Tonry}, {White}, {Ba{\~n}ados}, {Bell},
  {Bender}, {Bernard}, {Boegner}, {Boffi}, {Botticella}, {Calamida},
  {Casertano}, {Chen}, {Chen}, {Cole}, {Deacon}, {Frenk}, {Fitzsimmons},
  {Gezari}, {Gibbs}, {Goessl}, {Goggia}, {Gourgue}, {Goldman}, {Grant},
  {Grebel}, {Hambly}, {Hasinger}, {Heavens}, {Heckman}, {Henderson}, {Henning},
  {Holman}, {Hopp}, {Ip}, {Isani}, {Jackson}, {Keyes}, {Koekemoer}, {Kotak},
  {Le}, {Liska}, {Long}, {Lucey}, {Liu}, {Martin}, {Masci}, {McLean}, {Mindel},
  {Misra}, {Morganson}, {Murphy}, {Obaika}, {Narayan}, {Nieto-Santisteban},
  {Norberg}, {Peacock}, {Pier}, {Postman}, {Primak}, {Rae}, {Rai}, {Riess},
  {Riffeser}, {Rix}, {R{\"o}ser}, {Russel}, {Rutz}, {Schilbach}, {Schultz},
  {Scolnic}, {Strolger}, {Szalay}, {Seitz}, {Small}, {Smith}, {Soderblom},
  {Taylor}, {Thomson}, {Taylor}, {Thakar}, {Thiel}, {Thilker}, {Unger},
  {Urata}, {Valenti}, {Wagner}, {Walder}, {Walter}, {Watters}, {Werner},
  {Wood-Vasey}, \& {Wyse}}]{Chambers2016}
{Chambers}, K.~C., {Magnier}, E.~A., {Metcalfe}, N., {et~al.} 2016, arXiv
  e-prints, arXiv:1612.05560.
\newblock \doarXiv{1612.05560}

\bibitem[{{Chesley} {et~al.}(2016){Chesley}, {Farnocchia}, {Pravec}, \&
  {Vokrouhlick{\'y}}}]{Chesley2016}
{Chesley}, S.~R., {Farnocchia}, D., {Pravec}, P., \& {Vokrouhlick{\'y}}, D.
  2016, in IAU Symp. 318: Asteroids: New Observations, New Models, 250--258,
  \dodoi{10.1017/S1743921315008790}

\bibitem[{{Chesley} \& {Yeomans}(2005)}]{Chesley2005}
{Chesley}, S.~R., \& {Yeomans}, D.~K. 2005, in IAU Colloq. 197: Dynamics of
  Populations of Planetary Systems, 289--302, \dodoi{10.1017/S1743921304008786}

\bibitem[{{Christensen} {et~al.}(2016){Christensen}, {Carson Fuls}, {Gibbs},
  {Grauer}, {Johnson}, {Kowalski}, {Larson}, {Leonard}, {Matheny}, {Seaman}, \&
  {Shelly}}]{Christensen2016}
{Christensen}, E.~J., {Carson Fuls}, D., {Gibbs}, A., {et~al.} 2016, in
  AAS/Division for Planetary Sciences \#48, 405.01

\bibitem[{{Clark} {et~al.}(2009){Clark}, {Ockert-Bell}, {Cloutis}, {Nesvorny},
  {Moth{\'e}-Diniz}, \& {Bus}}]{Clark2009}
{Clark}, B.~E., {Ockert-Bell}, M.~E., {Cloutis}, E.~A., {et~al.} 2009, \icarus,
  202, 119, \dodoi{10.1016/j.icarus.2009.02.027}

\bibitem[{{Crovisier} {et~al.}(1995){Crovisier}, {Biver}, {Bockelee-Morvan},
  {Colom}, {Jorda}, {Lellouch}, {Paubert}, \& {Despois}}]{Crovisier1995}
{Crovisier}, J., {Biver}, N., {Bockelee-Morvan}, D., {et~al.} 1995, \icarus,
  115, 213, \dodoi{10.1006/icar.1995.1091}

\bibitem[{{Del Vigna} {et~al.}(2018){Del Vigna}, {Faggioli}, {Milani}, {Spoto},
  {Farnocchia}, \& {Carry}}]{DelVigna2018}
{Del Vigna}, A., {Faggioli}, L., {Milani}, A., {et~al.} 2018, \aap, 617, A61,
  \dodoi{10.1051/0004-6361/201833153}

\bibitem[{Desch \& Jackson(2021)}]{Desch20211i}
Desch, S.~J., \& Jackson, A.~P. 2021, Journal of Geophysical Research: Planets,
  e2020JE006807

\bibitem[{{Duncan} {et~al.}(2004){Duncan}, {Levison}, \& {Dones}}]{Duncan2004}
{Duncan}, M., {Levison}, H., \& {Dones}, L. 2004, in Comets II, 193--204

\bibitem[{{Eggl} {et~al.}(2020){Eggl}, {Farnocchia}, {Chamberlin}, \&
  {Chesley}}]{Eggl2020}
{Eggl}, S., {Farnocchia}, D., {Chamberlin}, A.~B., \& {Chesley}, S.~R. 2020,
  \icarus, 339, 113596, \dodoi{10.1016/j.icarus.2019.113596}

\bibitem[{{Farnocchia}(2021)}]{Farnocchia2021sb441}
{Farnocchia}, D. 2021, {Small-Body Perturber FilesSB441-N16 and SB441-N343},
  Tech. Rep. IOM 392R-21-005, Jet Propulsion Laboratory

\bibitem[{{Farnocchia} {et~al.}(2013){Farnocchia}, {Chesley}, {Chodas},
  {Micheli}, {Tholen}, {Milani}, {Elliott}, \&
  {Bernardi}}]{Farnocchia2013_yarko}
{Farnocchia}, D., {Chesley}, S.~R., {Chodas}, P.~W., {et~al.} 2013, \icarus,
  224, 192, \dodoi{10.1016/j.icarus.2013.02.020}

\bibitem[{{Farnocchia} {et~al.}(2014{\natexlab{a}}){Farnocchia}, {Chesley},
  {Chodas}, {Tricarico}, {Kelley}, \& {Farnham}}]{Farnocchia2014SidingSpring}
---. 2014{\natexlab{a}}, \apj, 790, 114, \dodoi{10.1088/0004-637X/790/2/114}

\bibitem[{{Farnocchia} {et~al.}(2015){Farnocchia}, {Chesley}, {Milani},
  {Gronchi}, \& {Chodas}}]{Farnocchia2015ast4}
{Farnocchia}, D., {Chesley}, S.~R., {Milani}, A., {Gronchi}, G.~F., \&
  {Chodas}, P.~W. 2015, in Asteroids IV, 815--834,
  \dodoi{10.2458/azu\_uapress\_9780816532131-ch041}

\bibitem[{{Farnocchia} {et~al.}(2014{\natexlab{b}}){Farnocchia}, {Chesley},
  {Tholen}, \& {Micheli}}]{Farnocchia2014Nyx}
{Farnocchia}, D., {Chesley}, S.~R., {Tholen}, D.~J., \& {Micheli}, M.
  2014{\natexlab{b}}, Celestial Mechanics and Dynamical Astronomy, 119, 301,
  \dodoi{10.1007/s10569-014-9536-9}

\bibitem[{{Farnocchia} {et~al.}(2017){Farnocchia}, {Tholen}, {Micheli}, {Ryan},
  {Rivera-Valentin}, {Taylor}, \& {Giorgini}}]{Farnocchia2017TC25}
{Farnocchia}, D., {Tholen}, D.~J., {Micheli}, M., {et~al.} 2017, in
  AAS/Division for Planetary Sciences Meeting Abstracts \#49, 100.09

\bibitem[{{Farnocchia} {et~al.}(2021){Farnocchia}, {Chesley}, {Takahashi},
  {Rozitis}, {Vokrouhlick{\'y}}, {Rush}, {Mastrodemos}, {Kennedy}, {Park},
  {Bellerose}, {Lubey}, {Velez}, {Davis}, {Emery}, {Leonard}, {Geeraert},
  {Antreasian}, \& {Lauretta}}]{Farnocchia2021Orex}
{Farnocchia}, D., {Chesley}, S.~R., {Takahashi}, Y., {et~al.} 2021, \icarus,
  369, 114594, \dodoi{10.1016/j.icarus.2021.114594}

\bibitem[{{Fedorets} {et~al.}(2020){Fedorets}, {Micheli}, {Jedicke}, {Naidu},
  {Farnocchia}, {Granvik}, {Moskovitz}, {Schwamb}, {Weryk}, {Wierzcho{\'s}},
  {Christensen}, {Pruyne}, {Bottke}, {Ye}, {Wainscoat}, {Devog{\`e}le},
  {Buchanan}, {Djupvik}, {Faes}, {F{\"o}hring}, {Roediger}, {Seccull}, \&
  {Smith}}]{Fedorets2020}
{Fedorets}, G., {Micheli}, M., {Jedicke}, R., {et~al.} 2020, \aj, 160, 277,
  \dodoi{10.3847/1538-3881/abc3bc}

\bibitem[{{Feng} \& {Jones}(2018)}]{Feng2018}
{Feng}, F., \& {Jones}, H.~R.~A. 2018, \apjl, 852, L27,
  \dodoi{10.3847/2041-8213/aaa404}

\bibitem[{{Fink}(2009)}]{Fink2009}
{Fink}, U. 2009, \icarus, 201, 311, \dodoi{10.1016/j.icarus.2008.12.044}

\bibitem[{{Gaia Collaboration} {et~al.}(2016){Gaia Collaboration}, {Prusti},
  {de Bruijne}, {Brown}, {Vallenari}, {Babusiaux}, {Bailer-Jones}, {Bastian},
  {Biermann}, {Evans}, {Eyer}, {Jansen}, {Jordi}, {Klioner}, {Lammers},
  {Lindegren}, {Luri}, {Mignard}, {Milligan}, {Panem}, {Poinsignon},
  {Pourbaix}, {Randich}, {Sarri}, {Sartoretti}, {Siddiqui}, {Soubiran},
  {Valette}, {van Leeuwen}, {Walton}, {Aerts}, {Arenou}, {Cropper}, {Drimmel},
  {H{\o}g}, {Katz}, {Lattanzi}, {O'Mullane}, {Grebel}, {Holland}, {Huc},
  {Passot}, {Bramante}, {Cacciari}, {Casta{\~n}eda}, {Chaoul}, {Cheek}, {De
  Angeli}, {Fabricius}, {Guerra}, {Hern{\'a}ndez}, {Jean-Antoine-Piccolo},
  {Masana}, {Messineo}, {Mowlavi}, {Nienartowicz}, {Ord{\'o}{\~n}ez-Blanco},
  {Panuzzo}, {Portell}, {Richards}, {Riello}, {Seabroke}, {Tanga},
  {Th{\'e}venin}, {Torra}, {Els}, {Gracia-Abril}, {Comoretto},
  {Garcia-Reinaldos}, {Lock}, {Mercier}, {Altmann}, {Andrae}, {Astraatmadja},
  {Bellas-Velidis}, {Benson}, {Berthier}, {Blomme}, {Busso}, {Carry},
  {Cellino}, {Clementini}, {Cowell}, {Creevey}, {Cuypers}, {Davidson}, {De
  Ridder}, {de Torres}, {Delchambre}, {Dell'Oro}, {Ducourant}, {Fr{\'e}mat},
  {Garc{\'\i}a-Torres}, {Gosset}, {Halbwachs}, {Hambly}, {Harrison}, {Hauser},
  {Hestroffer}, {Hodgkin}, {Huckle}, {Hutton}, {Jasniewicz}, {Jordan},
  {Kontizas}, {Korn}, {Lanzafame}, {Manteiga}, {Moitinho}, {Muinonen},
  {Osinde}, {Pancino}, {Pauwels}, {Petit}, {Recio-Blanco}, {Robin}, {Sarro},
  {Siopis}, {Smith}, {Smith}, {Sozzetti}, {Thuillot}, {van Reeven}, {Viala},
  {Abbas}, {Abreu Aramburu}, {Accart}, {Aguado}, {Allan}, {Allasia},
  {Altavilla}, {{\'A}lvarez}, {Alves}, {Anderson}, {Andrei}, {Anglada Varela},
  {Antiche}, {Antoja}, {Ant{\'o}n}, {Arcay}, {Atzei}, {Ayache}, {Bach},
  {Baker}, {Balaguer-N{\'u}{\~n}ez}, {Barache}, {Barata}, {Barbier}, {Barblan},
  {Baroni}, {Barrado y Navascu{\'e}s}, {Barros}, {Barstow}, {Becciani},
  {Bellazzini}, {Bellei}, {Bello Garc{\'\i}a}, {Belokurov}, {Bendjoya},
  {Berihuete}, {Bianchi}, {Bienaym{\'e}}, {Billebaud}, {Blagorodnova},
  {Blanco-Cuaresma}, {Boch}, {Bombrun}, {Borrachero}, {Bouquillon}, {Bourda},
  {Bouy}, {Bragaglia}, {Breddels}, {Brouillet}, {Br{\"u}semeister},
  {Bucciarelli}, {Budnik}, {Burgess}, {Burgon}, {Burlacu}, {Busonero}, {Buzzi},
  {Caffau}, {Cambras}, {Campbell}, {Cancelliere}, {Cantat-Gaudin}, {Carlucci},
  {Carrasco}, {Castellani}, {Charlot}, {Charnas}, {Charvet}, {Chassat},
  {Chiavassa}, {Clotet}, {Cocozza}, {Collins}, {Collins}, {Costigan}, {Crifo},
  {Cross}, {Crosta}, {Crowley}, {Dafonte}, {Damerdji}, {Dapergolas}, {David},
  {David}, {De Cat}, {de Felice}, {de Laverny}, {De Luise}, {De March}, {de
  Martino}, {de Souza}, {Debosscher}, {del Pozo}, {Delbo}, {Delgado},
  {Delgado}, {di Marco}, {Di Matteo}, {Diakite}, {Distefano}, {Dolding}, {Dos
  Anjos}, {Drazinos}, {Dur{\'a}n}, {Dzigan}, {Ecale}, {Edvardsson}, {Enke},
  {Erdmann}, {Escolar}, {Espina}, {Evans}, {Eynard Bontemps}, {Fabre},
  {Fabrizio}, {Faigler}, {Falc{\~a}o}, {Farr{\`a}s Casas}, {Faye}, {Federici},
  {Fedorets}, {Fern{\'a}ndez-Hern{\'a}ndez}, {Fernique}, {Fienga}, {Figueras},
  {Filippi}, {Findeisen}, {Fonti}, {Fouesneau}, {Fraile}, {Fraser}, {Fuchs},
  {Furnell}, {Gai}, {Galleti}, {Galluccio}, {Garabato}, {Garc{\'\i}a-Sedano},
  {Gar{\'e}}, {Garofalo}, {Garralda}, {Gavras}, {Gerssen}, {Geyer}, {Gilmore},
  {Girona}, {Giuffrida}, {Gomes}, {Gonz{\'a}lez-Marcos},
  {Gonz{\'a}lez-N{\'u}{\~n}ez}, {Gonz{\'a}lez-Vidal}, {Granvik}, {Guerrier},
  {Guillout}, {Guiraud}, {G{\'u}rpide}, {Guti{\'e}rrez-S{\'a}nchez}, {Guy},
  {Haigron}, {Hatzidimitriou}, {Haywood}, {Heiter}, {Helmi}, {Hobbs},
  {Hofmann}, {Holl}, {Holland}, {Hunt}, {Hypki}, {Icardi}, {Irwin}, {Jevardat
  de Fombelle}, {Jofr{\'e}}, {Jonker}, {Jorissen}, {Julbe}, {Karampelas},
  {Kochoska}, {Kohley}, {Kolenberg}, {Kontizas}, {Koposov}, {Kordopatis},
  {Koubsky}, {Kowalczyk}, {Krone-Martins}, {Kudryashova}, {Kull}, {Bachchan},
  {Lacoste-Seris}, {Lanza}, {Lavigne}, {Le Poncin-Lafitte}, {Lebreton},
  {Lebzelter}, {Leccia}, {Leclerc}, {Lecoeur-Taibi}, {Lemaitre}, {Lenhardt},
  {Leroux}, {Liao}, {Licata}, {Lindstr{\o}m}, {Lister}, {Livanou}, {Lobel},
  {L{\"o}ffler}, {L{\'o}pez}, {Lopez-Lozano}, {Lorenz}, {Loureiro},
  {MacDonald}, {Magalh{\~a}es Fernandes}, {Managau}, {Mann}, {Mantelet},
  {Marchal}, {Marchant}, {Marconi}, {Marie}, {Marinoni}, {Marrese},
  {Marschalk{\'o}}, {Marshall}, {Mart{\'\i}n-Fleitas}, {Martino}, {Mary},
  {Matijevi{\v{c}}}, {Mazeh}, {McMillan}, {Messina}, {Mestre}, {Michalik},
  {Millar}, {Miranda}, {Molina}, {Molinaro}, {Molinaro}, {Moln{\'a}r},
  {Moniez}, {Montegriffo}, {Monteiro}, {Mor}, {Mora}, {Morbidelli}, {Morel},
  {Morgenthaler}, {Morley}, {Morris}, {Mulone}, {Muraveva}, {Musella},
  {Narbonne}, {Nelemans}, {Nicastro}, {Noval}, {Ord{\'e}novic},
  {Ordieres-Mer{\'e}}, {Osborne}, {Pagani}, {Pagano}, {Pailler}, {Palacin},
  {Palaversa}, {Parsons}, {Paulsen}, {Pecoraro}, {Pedrosa}, {Pentik{\"a}inen},
  {Pereira}, {Pichon}, {Piersimoni}, {Pineau}, {Plachy}, {Plum}, {Poujoulet},
  {Pr{\v{s}}a}, {Pulone}, {Ragaini}, {Rago}, {Rambaux}, {Ramos-Lerate},
  {Ranalli}, {Rauw}, {Read}, {Regibo}, {Renk}, {Reyl{\'e}}, {Ribeiro},
  {Rimoldini}, {Ripepi}, {Riva}, {Rixon}, {Roelens}, {Romero-G{\'o}mez},
  {Rowell}, {Royer}, {Rudolph}, {Ruiz-Dern}, {Sadowski}, {Sagrist{\`a}
  Sell{\'e}s}, {Sahlmann}, {Salgado}, {Salguero}, {Sarasso}, {Savietto},
  {Schnorhk}, {Schultheis}, {Sciacca}, {Segol}, {Segovia}, {Segransan},
  {Serpell}, {Shih}, {Smareglia}, {Smart}, {Smith}, {Solano}, {Solitro},
  {Sordo}, {Soria Nieto}, {Souchay}, {Spagna}, {Spoto}, {Stampa}, {Steele},
  {Steidelm{\"u}ller}, {Stephenson}, {Stoev}, {Suess}, {S{\"u}veges}, {Surdej},
  {Szabados}, {Szegedi-Elek}, {Tapiador}, {Taris}, {Tauran}, {Taylor},
  {Teixeira}, {Terrett}, {Tingley}, {Trager}, {Turon}, {Ulla}, {Utrilla},
  {Valentini}, {van Elteren}, {Van Hemelryck}, {van Leeuwen}, {Varadi},
  {Vecchiato}, {Veljanoski}, {Via}, {Vicente}, {Vogt}, {Voss}, {Votruba},
  {Voutsinas}, {Walmsley}, {Weiler}, {Weingrill}, {Werner}, {Wevers},
  {Whitehead}, {Wyrzykowski}, {Yoldas}, {{\v{Z}}erjal}, {Zucker}, {Zurbach},
  {Zwitter}, {Alecu}, {Allen}, {Allende Prieto}, {Amorim},
  {Anglada-Escud{\'e}}, {Arsenijevic}, {Azaz}, {Balm}, {Beck}, {Bernstein},
  {Bigot}, {Bijaoui}, {Blasco}, {Bonfigli}, {Bono}, {Boudreault}, {Bressan},
  {Brown}, {Brunet}, {Bunclark}, {Buonanno}, {Butkevich}, {Carret}, {Carrion},
  {Chemin}, {Ch{\'e}reau}, {Corcione}, {Darmigny}, {de Boer}, {de Teodoro}, {de
  Zeeuw}, {Delle Luche}, {Domingues}, {Dubath}, {Fodor}, {Fr{\'e}zouls},
  {Fries}, {Fustes}, {Fyfe}, {Gallardo}, {Gallegos}, {Gardiol}, {Gebran},
  {Gomboc}, {G{\'o}mez}, {Grux}, {Gueguen}, {Heyrovsky}, {Hoar}, {Iannicola},
  {Isasi Parache}, {Janotto}, {Joliet}, {Jonckheere}, {Keil}, {Kim},
  {Klagyivik}, {Klar}, {Knude}, {Kochukhov}, {Kolka}, {Kos}, {Kutka}, {Lainey},
  {LeBouquin}, {Liu}, {Loreggia}, {Makarov}, {Marseille}, {Martayan},
  {Martinez-Rubi}, {Massart}, {Meynadier}, {Mignot}, {Munari}, {Nguyen},
  {Nordlander}, {Ocvirk}, {O'Flaherty}, {Olias Sanz}, {Ortiz}, {Osorio},
  {Oszkiewicz}, {Ouzounis}, {Palmer}, {Park}, {Pasquato}, {Peltzer}, {Peralta},
  {P{\'e}turaud}, {Pieniluoma}, {Pigozzi}, {Poels}, {Prat}, {Prod'homme},
  {Raison}, {Rebordao}, {Risquez}, {Rocca-Volmerange}, {Rosen}, {Ruiz-Fuertes},
  {Russo}, {Sembay}, {Serraller Vizcaino}, {Short}, {Siebert}, {Silva},
  {Sinachopoulos}, {Slezak}, {Soffel}, {Sosnowska}, {Strai{\v{z}}ys}, {ter
  Linden}, {Terrell}, {Theil}, {Tiede}, {Troisi}, {Tsalmantza}, {Tur},
  {Vaccari}, {Vachier}, {Valles}, {Van Hamme}, {Veltz}, {Virtanen}, {Wallut},
  {Wichmann}, {Wilkinson}, {Ziaeepour}, \& {Zschocke}}]{Gaia16}
{Gaia Collaboration}, {Prusti}, T., {de Bruijne}, J.~H.~J., {et~al.} 2016,
  \aap, 595, A1, \dodoi{10.1051/0004-6361/201629272}

\bibitem[{{Gaia Collaboration} {et~al.}(2018){Gaia Collaboration}, {Brown},
  {Vallenari}, {Prusti}, {de Bruijne}, {Babusiaux}, {Bailer-Jones}, {Biermann},
  {Evans}, {Eyer}, {Jansen}, {Jordi}, {Klioner}, {Lammers}, {Lindegren},
  {Luri}, {Mignard}, {Panem}, {Pourbaix}, {Randich}, {Sartoretti}, {Siddiqui},
  {Soubiran}, {van Leeuwen}, {Walton}, {Arenou}, {Bastian}, {Cropper},
  {Drimmel}, {Katz}, {Lattanzi}, {Bakker}, {Cacciari}, {Casta{\~n}eda},
  {Chaoul}, {Cheek}, {De Angeli}, {Fabricius}, {Guerra}, {Holl}, {Masana},
  {Messineo}, {Mowlavi}, {Nienartowicz}, {Panuzzo}, {Portell}, {Riello},
  {Seabroke}, {Tanga}, {Th{\'e}venin}, {Gracia-Abril}, {Comoretto},
  {Garcia-Reinaldos}, {Teyssier}, {Altmann}, {Andrae}, {Audard},
  {Bellas-Velidis}, {Benson}, {Berthier}, {Blomme}, {Burgess}, {Busso},
  {Carry}, {Cellino}, {Clementini}, {Clotet}, {Creevey}, {Davidson}, {De
  Ridder}, {Delchambre}, {Dell'Oro}, {Ducourant},
  {Fern{\'a}ndez-Hern{\'a}ndez}, {Fouesneau}, {Fr{\'e}mat}, {Galluccio},
  {Garc{\'\i}a-Torres}, {Gonz{\'a}lez-N{\'u}{\~n}ez}, {Gonz{\'a}lez-Vidal},
  {Gosset}, {Guy}, {Halbwachs}, {Hambly}, {Harrison}, {Hern{\'a}ndez},
  {Hestroffer}, {Hodgkin}, {Hutton}, {Jasniewicz}, {Jean-Antoine-Piccolo},
  {Jordan}, {Korn}, {Krone-Martins}, {Lanzafame}, {Lebzelter}, {L{\"o}ffler},
  {Manteiga}, {Marrese}, {Mart{\'\i}n-Fleitas}, {Moitinho}, {Mora}, {Muinonen},
  {Osinde}, {Pancino}, {Pauwels}, {Petit}, {Recio-Blanco}, {Richards},
  {Rimoldini}, {Robin}, {Sarro}, {Siopis}, {Smith}, {Sozzetti}, {S{\"u}veges},
  {Torra}, {van Reeven}, {Abbas}, {Abreu Aramburu}, {Accart}, {Aerts},
  {Altavilla}, {{\'A}lvarez}, {Alvarez}, {Alves}, {Anderson}, {Andrei},
  {Anglada Varela}, {Antiche}, {Antoja}, {Arcay}, {Astraatmadja}, {Bach},
  {Baker}, {Balaguer-N{\'u}{\~n}ez}, {Balm}, {Barache}, {Barata}, {Barbato},
  {Barblan}, {Barklem}, {Barrado}, {Barros}, {Barstow}, {Bartholom{\'e}
  Mu{\~n}oz}, {Bassilana}, {Becciani}, {Bellazzini}, {Berihuete}, {Bertone},
  {Bianchi}, {Bienaym{\'e}}, {Blanco-Cuaresma}, {Boch}, {Boeche}, {Bombrun},
  {Borrachero}, {Bossini}, {Bouquillon}, {Bourda}, {Bragaglia}, {Bramante},
  {Breddels}, {Bressan}, {Brouillet}, {Br{\"u}semeister}, {Brugaletta},
  {Bucciarelli}, {Burlacu}, {Busonero}, {Butkevich}, {Buzzi}, {Caffau},
  {Cancelliere}, {Cannizzaro}, {Cantat-Gaudin}, {Carballo}, {Carlucci},
  {Carrasco}, {Casamiquela}, {Castellani}, {Castro-Ginard}, {Charlot},
  {Chemin}, {Chiavassa}, {Cocozza}, {Costigan}, {Cowell}, {Crifo}, {Crosta},
  {Crowley}, {Cuypers}, {Dafonte}, {Damerdji}, {Dapergolas}, {David}, {David},
  {de Laverny}, {De Luise}, {De March}, {de Martino}, {de Souza}, {de Torres},
  {Debosscher}, {del Pozo}, {Delbo}, {Delgado}, {Delgado}, {Di Matteo},
  {Diakite}, {Diener}, {Distefano}, {Dolding}, {Drazinos}, {Dur{\'a}n},
  {Edvardsson}, {Enke}, {Eriksson}, {Esquej}, {Eynard Bontemps}, {Fabre},
  {Fabrizio}, {Faigler}, {Falc{\~a}o}, {Farr{\`a}s Casas}, {Federici},
  {Fedorets}, {Fernique}, {Figueras}, {Filippi}, {Findeisen}, {Fonti},
  {Fraile}, {Fraser}, {Fr{\'e}zouls}, {Gai}, {Galleti}, {Garabato},
  {Garc{\'\i}a-Sedano}, {Garofalo}, {Garralda}, {Gavel}, {Gavras}, {Gerssen},
  {Geyer}, {Giacobbe}, {Gilmore}, {Girona}, {Giuffrida}, {Glass}, {Gomes},
  {Granvik}, {Gueguen}, {Guerrier}, {Guiraud}, {Guti{\'e}rrez-S{\'a}nchez},
  {Haigron}, {Hatzidimitriou}, {Hauser}, {Haywood}, {Heiter}, {Helmi}, {Heu},
  {Hilger}, {Hobbs}, {Hofmann}, {Holland}, {Huckle}, {Hypki}, {Icardi},
  {Jan{\ss}en}, {Jevardat de Fombelle}, {Jonker}, {Juh{\'a}sz}, {Julbe},
  {Karampelas}, {Kewley}, {Klar}, {Kochoska}, {Kohley}, {Kolenberg},
  {Kontizas}, {Kontizas}, {Koposov}, {Kordopatis}, {Kostrzewa-Rutkowska},
  {Koubsky}, {Lambert}, {Lanza}, {Lasne}, {Lavigne}, {Le Fustec}, {Le
  Poncin-Lafitte}, {Lebreton}, {Leccia}, {Leclerc}, {Lecoeur-Taibi},
  {Lenhardt}, {Leroux}, {Liao}, {Licata}, {Lindstr{\o}m}, {Lister}, {Livanou},
  {Lobel}, {L{\'o}pez}, {Managau}, {Mann}, {Mantelet}, {Marchal}, {Marchant},
  {Marconi}, {Marinoni}, {Marschalk{\'o}}, {Marshall}, {Martino}, {Marton},
  {Mary}, {Massari}, {Matijevi{\v{c}}}, {Mazeh}, {McMillan}, {Messina},
  {Michalik}, {Millar}, {Molina}, {Molinaro}, {Moln{\'a}r}, {Montegriffo},
  {Mor}, {Morbidelli}, {Morel}, {Morris}, {Mulone}, {Muraveva}, {Musella},
  {Nelemans}, {Nicastro}, {Noval}, {O'Mullane}, {Ord{\'e}novic},
  {Ord{\'o}{\~n}ez-Blanco}, {Osborne}, {Pagani}, {Pagano}, {Pailler},
  {Palacin}, {Palaversa}, {Panahi}, {Pawlak}, {Piersimoni}, {Pineau}, {Plachy},
  {Plum}, {Poggio}, {Poujoulet}, {Pr{\v{s}}a}, {Pulone}, {Racero}, {Ragaini},
  {Rambaux}, {Ramos-Lerate}, {Regibo}, {Reyl{\'e}}, {Riclet}, {Ripepi}, {Riva},
  {Rivard}, {Rixon}, {Roegiers}, {Roelens}, {Romero-G{\'o}mez}, {Rowell},
  {Royer}, {Ruiz-Dern}, {Sadowski}, {Sagrist{\`a} Sell{\'e}s}, {Sahlmann},
  {Salgado}, {Salguero}, {Sanna}, {Santana-Ros}, {Sarasso}, {Savietto},
  {Schultheis}, {Sciacca}, {Segol}, {Segovia}, {S{\'e}gransan}, {Shih},
  {Siltala}, {Silva}, {Smart}, {Smith}, {Solano}, {Solitro}, {Sordo}, {Soria
  Nieto}, {Souchay}, {Spagna}, {Spoto}, {Stampa}, {Steele},
  {Steidelm{\"u}ller}, {Stephenson}, {Stoev}, {Suess}, {Surdej}, {Szabados},
  {Szegedi-Elek}, {Tapiador}, {Taris}, {Tauran}, {Taylor}, {Teixeira},
  {Terrett}, {Teyssandier}, {Thuillot}, {Titarenko}, {Torra Clotet}, {Turon},
  {Ulla}, {Utrilla}, {Uzzi}, {Vaillant}, {Valentini}, {Valette}, {van Elteren},
  {Van Hemelryck}, {van Leeuwen}, {Vaschetto}, {Vecchiato}, {Veljanoski},
  {Viala}, {Vicente}, {Vogt}, {von Essen}, {Voss}, {Votruba}, {Voutsinas},
  {Walmsley}, {Weiler}, {Wertz}, {Wevers}, {Wyrzykowski}, {Yoldas},
  {{\v{Z}}erjal}, {Ziaeepour}, {Zorec}, {Zschocke}, {Zucker}, {Zurbach}, \&
  {Zwitter}}]{Gaia18}
{Gaia Collaboration}, {Brown}, A.~G.~A., {Vallenari}, A., {et~al.} 2018, \aap,
  616, A1, \dodoi{10.1051/0004-6361/201833051}

\bibitem[{{Gehrels} \& {Taylor}(1977)}]{Gehrels1977}
{Gehrels}, T., \& {Taylor}, R.~C. 1977, \aj, 82, 229, \dodoi{10.1086/112036}

\bibitem[{{Granvik} {et~al.}(2018){Granvik}, {Morbidelli}, {Jedicke}, {Bolin},
  {Bottke}, {Beshore}, {Vokrouhlick{\'y}}, {Nesvorn{\'y}}, \&
  {Michel}}]{Granvik2018}
{Granvik}, M., {Morbidelli}, A., {Jedicke}, R., {et~al.} 2018, \icarus, 312,
  181, \dodoi{10.1016/j.icarus.2018.04.018}

\bibitem[{{Greenberg} {et~al.}(2020){Greenberg}, {Margot}, {Verma}, {Taylor},
  \& {Hodge}}]{Greenberg2020}
{Greenberg}, A.~H., {Margot}, J.-L., {Verma}, A.~K., {Taylor}, P.~A., \&
  {Hodge}, S.~E. 2020, \aj, 159, 92, \dodoi{10.3847/1538-3881/ab62a3}

\bibitem[{{Gunnarsson} {et~al.}(2008){Gunnarsson}, {Bockel{\'e}e-Morvan},
  {Biver}, {Crovisier}, \& {Rickman}}]{Gunnarsson2008}
{Gunnarsson}, M., {Bockel{\'e}e-Morvan}, D., {Biver}, N., {Crovisier}, J., \&
  {Rickman}, H. 2008, \aap, 484, 537, \dodoi{10.1051/0004-6361:20078069}

\bibitem[{{Hainaut} {et~al.}(2021){Hainaut}, {Micheli}, {Cano}, {Mart{\'\i}n},
  {Faggioli}, \& {Cennamo}}]{Hainaut+21}
{Hainaut}, O.~R., {Micheli}, M., {Cano}, J.~L., {et~al.} 2021, \aap, 653, A124,
  \dodoi{10.1051/0004-6361/202141519}

\bibitem[{{Hsieh} {et~al.}(2020){Hsieh}, {Novakovi{\'c}}, {Walsh}, \&
  {Sch{\"o}rghofer}}]{Hsieh2020}
{Hsieh}, H.~H., {Novakovi{\'c}}, B., {Walsh}, K.~J., \& {Sch{\"o}rghofer}, N.
  2020, \aj, 159, 179, \dodoi{10.3847/1538-3881/ab7899}

\bibitem[{{Ieva} {et~al.}(2020){Ieva}, {Dotto}, {Mazzotta Epifani}, {Perna},
  {Fanasca}, {Lazzarin}, {Bertini}, {Petropoulou}, {Rossi}, {Micheli}, \&
  {Perozzi}}]{Ieva2020}
{Ieva}, S., {Dotto}, E., {Mazzotta Epifani}, E., {et~al.} 2020, \aap, 644, A23,
  \dodoi{10.1051/0004-6361/202038968}

\bibitem[{Jackson \& Desch(2021)}]{Jackson20211i}
Jackson, A.~P., \& Desch, S.~J. 2021, Journal of Geophysical Research: Planets,
  e2020JE006706

\bibitem[{{Jedicke} {et~al.}(2015){Jedicke}, {Granvik}, {Micheli}, {Ryan},
  {Spahr}, \& {Yeomans}}]{Jedicke2015}
{Jedicke}, R., {Granvik}, M., {Micheli}, M., {et~al.} 2015, in Asteroids IV,
  795--813, \dodoi{10.2458/azu\_uapress\_9780816532131-ch040}

\bibitem[{{Jewitt} \& {Seligman}(2022)}]{Jewitt2022}
{Jewitt}, D., \& {Seligman}, D.~Z. 2022, arXiv e-prints, arXiv:2209.08182.
\newblock \doarXiv{2209.08182}

\bibitem[{{Jordi} {et~al.}(2010){Jordi}, {Gebran}, {Carrasco}, {de Bruijne},
  {Voss}, {Fabricius}, {Knude}, {Vallenari}, {Kohley}, \& {Mora}}]{Jordi2010}
{Jordi}, C., {Gebran}, M., {Carrasco}, J.~M., {et~al.} 2010, \aap, 523, A48,
  \dodoi{10.1051/0004-6361/201015441}

\bibitem[{{Kr{\'o}likowska}(2004)}]{Krolikowska2004}
{Kr{\'o}likowska}, M. 2004, \aap, 427, 1117, \dodoi{10.1051/0004-6361:20041339}

\bibitem[{{Lagerkvist} \& {Lagerros}(1997)}]{Lagerkvist1997}
{Lagerkvist}, C.~I., \& {Lagerros}, J.~S.~V. 1997, Astronomische Nachrichten,
  318, 391, \dodoi{10.1002/asna.2113180611}

\bibitem[{{Marschall} {et~al.}(2020){Marschall}, {Markkanen}, {Gerig},
  {Pinz{\'o}n-Rodr{\'\i}guez}, {Thomas}, \& {Wu}}]{Marschall2020}
{Marschall}, R., {Markkanen}, J., {Gerig}, S.-B., {et~al.} 2020, Frontiers in
  Physics, 8, 227, \dodoi{10.3389/fphy.2020.00227}

\bibitem[{{Marsden} {et~al.}(1973){Marsden}, {Sekanina}, \&
  {Yeomans}}]{Marsden1973}
{Marsden}, B.~G., {Sekanina}, Z., \& {Yeomans}, D.~K. 1973, \aj, 78, 211,
  \dodoi{10.1086/111402}

\bibitem[{{McMillan} {et~al.}(2016){McMillan}, {Larsen}, {Bressi}, {Scotti},
  {Mastaler}, \& {Tubbiolo}}]{McMillan2016}
{McMillan}, R.~S., {Larsen}, J.~A., {Bressi}, T.~H., {et~al.} 2016, in IAU
  Symp. 318: Asteroids: New Observations, New Models, 317--318,
  \dodoi{10.1017/S1743921315006766}

\bibitem[{{Meech} \& {Belton}(1990)}]{Meech1990}
{Meech}, K.~J., \& {Belton}, M. J.~S. 1990, \aj, 100, 1323,
  \dodoi{10.1086/115600}

\bibitem[{{Meech} {et~al.}(1997){Meech}, {Buie}, {Samarasinha}, {Mueller}, \&
  {Belton}}]{Meech1997}
{Meech}, K.~J., {Buie}, M.~W., {Samarasinha}, N.~H., {Mueller}, B. E.~A., \&
  {Belton}, M. J.~S. 1997, \aj, 113, 844, \dodoi{10.1086/118305}

\bibitem[{{Meech} {et~al.}(2001){Meech}, {Fern{\'a}ndez}, \&
  {Pittichov{\'a}}}]{meech2001}
{Meech}, K.~J., {Fern{\'a}ndez}, Y., \& {Pittichov{\'a}}, J. 2001, in
  AAS/Division for Planetary Sciences Meeting Abstracts, Vol.~33, AAS/Division
  for Planetary Sciences Meeting Abstracts \#33, 20.06

\bibitem[{{Meech} \& {Jewitt}(1987)}]{Meech1987}
{Meech}, K.~J., \& {Jewitt}, D.~C. 1987, \aap, 187, 585

\bibitem[{{Meech} \& {Svoren}(2004)}]{Meech2004}
{Meech}, K.~J., \& {Svoren}, J. 2004, in Comets II, 317--335

\bibitem[{{Meech} {et~al.}(2017){Meech}, {Schambeau}, {Sorli}, {Kleyna},
  {Micheli}, {Bauer}, {Denneau}, {Keane}, {Toller}, {Wainscoat}, {Hainaut},
  {Bhatt}, {Sahu}, {Yang}, {Kramer}, \& {Magnier}}]{Meech2017}
{Meech}, K.~J., {Schambeau}, C.~A., {Sorli}, K., {et~al.} 2017, \aj, 153, 206,
  \dodoi{10.3847/1538-3881/aa63f2}

\bibitem[{{Micheli} {et~al.}(2012){Micheli}, {Tholen}, \&
  {Elliott}}]{Micheli2012}
{Micheli}, M., {Tholen}, D.~J., \& {Elliott}, G.~T. 2012, \na, 17, 446,
  \dodoi{10.1016/j.newast.2011.11.008}

\bibitem[{{Micheli} {et~al.}(2013){Micheli}, {Tholen}, \&
  {Elliott}}]{Micheli2013}
---. 2013, \icarus, 226, 251, \dodoi{10.1016/j.icarus.2013.05.032}

\bibitem[{{Micheli} {et~al.}(2014){Micheli}, {Tholen}, \&
  {Elliott}}]{Micheli2014}
---. 2014, \apjl, 788, L1, \dodoi{10.1088/2041-8205/788/1/L1}

\bibitem[{{Micheli} {et~al.}(2018){Micheli}, {Farnocchia}, {Meech}, {Buie},
  {Hainaut}, {Prialnik}, {Sch{\"o}rghofer}, {Weaver}, {Chodas}, {Kleyna},
  {Weryk}, {Wainscoat}, {Ebeling}, {Keane}, {Chambers}, {Koschny}, \&
  {Petropoulos}}]{Micheli2018}
{Micheli}, M., {Farnocchia}, D., {Meech}, K.~J., {et~al.} 2018, \nat, 559, 223,
  \dodoi{10.1038/s41586-018-0254-4}

\bibitem[{{Mommert} {et~al.}(2014{\natexlab{a}}){Mommert}, {Hora},
  {Farnocchia}, {Chesley}, {Vokrouhlick{\'y}}, {Trilling}, {Mueller}, {Harris},
  {Smith}, \& {Fazio}}]{Mommert2014bd}
{Mommert}, M., {Hora}, J.~L., {Farnocchia}, D., {et~al.} 2014{\natexlab{a}},
  \apj, 786, 148, \dodoi{10.1088/0004-637X/786/2/148}

\bibitem[{{Mommert} {et~al.}(2014{\natexlab{b}}){Mommert}, {Farnocchia},
  {Hora}, {Chesley}, {Trilling}, {Chodas}, {Mueller}, {Harris}, {Smith}, \&
  {Fazio}}]{Mommert2014md}
{Mommert}, M., {Farnocchia}, D., {Hora}, J.~L., {et~al.} 2014{\natexlab{b}},
  \apjl, 789, L22, \dodoi{10.1088/2041-8205/789/1/L22}

\bibitem[{{Ostro} {et~al.}(2002){Ostro}, {Hudson}, {Benner}, {Giorgini},
  {Magri}, {Margot}, \& {Nolan}}]{Ostro2002}
{Ostro}, S.~J., {Hudson}, R.~S., {Benner}, L.~A.~M., {et~al.} 2002, in
  Asteroids III, 151--168

\bibitem[{{Paganini} {et~al.}(2013){Paganini}, {Mumma}, {Boehnhardt},
  {DiSanti}, {Villanueva}, {Bonev}, {Lippi}, {K{\"a}ufl}, \&
  {Blake}}]{Paganini2013}
{Paganini}, L., {Mumma}, M.~J., {Boehnhardt}, H., {et~al.} 2013, \apj, 766,
  100, \dodoi{10.1088/0004-637X/766/2/100}

\bibitem[{{Park} {et~al.}(2021){Park}, {Folkner}, {Williams}, \&
  {Boggs}}]{Park2021}
{Park}, R.~S., {Folkner}, W.~M., {Williams}, J.~G., \& {Boggs}, D.~H. 2021,
  \aj, 161, 105, \dodoi{10.3847/1538-3881/abd414}

\bibitem[{{Pravdo} {et~al.}(1999){Pravdo}, {Rabinowitz}, {Helin}, {Lawrence},
  {Bambery}, {Clark}, {Groom}, {Levin}, {Lorre}, {Shaklan}, {Kervin},
  {Africano}, {Sydney}, \& {Soohoo}}]{Pravdo1999}
{Pravdo}, S.~H., {Rabinowitz}, D.~L., {Helin}, E.~F., {et~al.} 1999, \aj, 117,
  1616, \dodoi{10.1086/300769}

\bibitem[{{Rafikov}(2018)}]{Rafikov2018}
{Rafikov}, R.~R. 2018, \apjl, 867, L17, \dodoi{10.3847/2041-8213/aae977}

\bibitem[{{Reach} {et~al.}(2013){Reach}, {Kelley}, \& {Vaubaillon}}]{reach2013}
{Reach}, W.~T., {Kelley}, M.~S., \& {Vaubaillon}, J. 2013, \icarus, 226, 777,
  \dodoi{10.1016/j.icarus.2013.06.011}

\bibitem[{{Seidelmann}(1977)}]{Seidelmann77}
{Seidelmann}, P.~K. 1977, Celestial Mechanics, 16, 165,
  \dodoi{10.1007/BF01228598}

\bibitem[{{Seligman} \& {Laughlin}(2020)}]{Seligman2020}
{Seligman}, D., \& {Laughlin}, G. 2020, \apjl, 896, L8,
  \dodoi{10.3847/2041-8213/ab963f}

\bibitem[{{Seligman} {et~al.}(2021){Seligman}, {Levine}, {Cabot}, {Laughlin},
  \& {Meech}}]{Seligman2021}
{Seligman}, D.~Z., {Levine}, W.~G., {Cabot}, S. H.~C., {Laughlin}, G., \&
  {Meech}, K. 2021, \apj, 920, 28, \dodoi{10.3847/1538-4357/ac1594}

\bibitem[{{Senay} \& {Jewitt}(1994)}]{Senay1994}
{Senay}, M.~C., \& {Jewitt}, D. 1994, \nat, 371, 229, \dodoi{10.1038/371229a0}

\bibitem[{{Slemp} {et~al.}(2022){Slemp}, {Meech}, {Bufanda}, {Kleyna},
  {Hainaut}, {Bauer}, {Weryk}, {Denneau}, {Keane}, {Bhatt}, {Sahu}, {Urasaki},
  {Wainscoat}, \& {Micheli}}]{Slemp2022}
{Slemp}, L.~A., {Meech}, K.~J., {Bufanda}, E., {et~al.} 2022, \psj, 3, 34,
  \dodoi{10.3847/PSJ/ac480d}

\bibitem[{{Snodgrass} {et~al.}(2017){Snodgrass}, {Agarwal}, {Combi},
  {Fitzsimmons}, {Guilbert-Lepoutre}, {Hsieh}, {Hui}, {Jehin}, {Kelley},
  {Knight}, {Opitom}, {Orosei}, {de Val-Borro}, \& {Yang}}]{Snodgrass2017}
{Snodgrass}, C., {Agarwal}, J., {Combi}, M., {et~al.} 2017, \aapr, 25, 5,
  \dodoi{10.1007/s00159-017-0104-7}

\bibitem[{{Spohn} {et~al.}(2015){Spohn}, {Knollenberg}, {Ball},
  {Banaszkiewicz}, {Benkhoff}, {Grott}, {Grygorczuk}, {H{\"u}ttig},
  {Hagermann}, {Kargl}, {Kaufmann}, {K{\"o}mle}, {K{\"u}hrt}, {Kossacki},
  {Marczewski}, {Pelivan}, {Schr{\"o}dter}, \& {Seiferlin}}]{Spohn2015}
{Spohn}, T., {Knollenberg}, J., {Ball}, A.~J., {et~al.} 2015, Science, 349,
  2.464, \dodoi{10.1126/science.aab0464}

\bibitem[{{Taylor} {et~al.}(2022){Taylor}, {Seligman}, {MacAyeal}, {Hainaut},
  \& {Meech}}]{Taylor2022}
{Taylor}, A.~G., {Seligman}, D.~Z., {MacAyeal}, D.~R., {Hainaut}, O.~R., \&
  {Meech}, K.~J. 2022, arXiv e-prints, arXiv:2209.15074.
\newblock \doarXiv{2209.15074}

\bibitem[{{Toliou} {et~al.}(2021){Toliou}, {Granvik}, \&
  {Tsirvoulis}}]{Toliou2021}
{Toliou}, A., {Granvik}, M., \& {Tsirvoulis}, G. 2021, \mnras, 506, 3301,
  \dodoi{10.1093/mnras/stab1934}

\bibitem[{{Trilling} {et~al.}(2018){Trilling}, {Mommert}, {Hora}, {Farnocchia},
  {Chodas}, {Giorgini}, {Smith}, {Carey}, {Lisse}, {Werner}, {McNeill},
  {Chesley}, {Emery}, {Fazio}, {Fernandez}, {Harris}, {Marengo}, {Mueller},
  {Roegge}, {Smith}, {Weaver}, {Meech}, \& {Micheli}}]{Trilling2018}
{Trilling}, D.~E., {Mommert}, M., {Hora}, J.~L., {et~al.} 2018, \aj, 156, 261,
  \dodoi{10.3847/1538-3881/aae88f}

\bibitem[{{Vere{\v{s}}} {et~al.}(2017){Vere{\v{s}}}, {Farnocchia}, {Chesley},
  \& {Chamberlin}}]{Veres2017}
{Vere{\v{s}}}, P., {Farnocchia}, D., {Chesley}, S.~R., \& {Chamberlin}, A.~B.
  2017, \icarus, 296, 139, \dodoi{10.1016/j.icarus.2017.05.021}

\bibitem[{{Vere{\v{s}}} {et~al.}(2012){Vere{\v{s}}}, {Jedicke}, {Denneau},
  {Wainscoat}, {Holman}, \& {Lin}}]{Veres2012}
{Vere{\v{s}}}, P., {Jedicke}, R., {Denneau}, L., {et~al.} 2012, \pasp, 124,
  1197, \dodoi{10.1086/668616}

\bibitem[{{Vokrouhlick{\'y}} {et~al.}(2015){Vokrouhlick{\'y}}, {Bottke},
  {Chesley}, {Scheeres}, \& {Statler}}]{Vokrouhlicky2015_ast4}
{Vokrouhlick{\'y}}, D., {Bottke}, W.~F., {Chesley}, S.~R., {Scheeres}, D.~J.,
  \& {Statler}, T.~S. 2015, in Asteroids IV, 509--531,
  \dodoi{10.2458/azu\_uapress\_9780816532131-ch027}

\bibitem[{{Vokrouhlick{\'y}} \& {Milani}(2000)}]{Vokrouhlicky2000}
{Vokrouhlick{\'y}}, D., \& {Milani}, A. 2000, \aap, 362, 746

\bibitem[{{Wainscoat} {et~al.}(2016){Wainscoat}, {Chambers}, {Lilly}, {Weryk},
  {Chastel}, {Denneau}, \& {Micheli}}]{Wainscoat2016}
{Wainscoat}, R., {Chambers}, K., {Lilly}, E., {et~al.} 2016, in IAU Symp. 318:
  Asteroids: New Observations, New Models, 293--298,
  \dodoi{10.1017/S1743921315009187}

\bibitem[{{Wainscoat} {et~al.}(2020){Wainscoat}, {Weryk}, {Ramanjooloo},
  {Chastel}, {Huber}, {Chambers}, \& {Magnier}}]{Wainscoat2020}
{Wainscoat}, R., {Weryk}, R., {Ramanjooloo}, Y., {et~al.} 2020, in AAS/Division
  for Planetary Sciences \#52, 107.03

\bibitem[{{Wierzchos} \& {Womack}(2020)}]{Wierzchos2020}
{Wierzchos}, K., \& {Womack}, M. 2020, \aj, 159, 136,
  \dodoi{10.3847/1538-3881/ab6e68}

\bibitem[{{Ye} {et~al.}(2016){Ye}, {Zhao}, \& {Li}}]{Ye2016}
{Ye}, J.-h., {Zhao}, H.-b., \& {Li}, B. 2016, \caa, 40, 54,
  \dodoi{10.1016/j.chinastron.2016.01.006}

\bibitem[{{Yeomans} \& {Chodas}(1989)}]{Yeomans89}
{Yeomans}, D.~K., \& {Chodas}, P.~W. 1989, \aj, 98, 1083,
  \dodoi{10.1086/115198}

\bibitem[{{Yeomans} {et~al.}(2004){Yeomans}, {Chodas}, {Sitarski}, {Szutowicz},
  \& {Kr{\'o}likowska}}]{Yeomans2004}
{Yeomans}, D.~K., {Chodas}, P.~W., {Sitarski}, G., {Szutowicz}, S., \&
  {Kr{\'o}likowska}, M. 2004, in Comets II, 137--151

\end{thebibliography}
\bibliographystyle{aasjournal}



\end{document}